\documentclass[journal,a4paper]{IEEEtran}

\usepackage{cite}
\usepackage{array}
\usepackage{fixltx2e}
\usepackage{color}
\usepackage{psfrag}
\usepackage{epsfig}
\usepackage{tabularx}
\usepackage{amsmath}
\usepackage{amssymb}
\usepackage{amsfonts}
\usepackage{algorithm,algorithmic}
\usepackage{pstricks, pst-node, pst-plot, pst-circ}
\usepackage{moredefs}

\def\ve#1{{\mathchoice{\mbox{\boldmath$\displaystyle #1$}}%
		      {\mbox{\boldmath$\textstyle #1$}}%
		      {\mbox{\boldmath$\scriptstyle #1$}}%
		      {\mbox{\boldmath$\scriptscriptstyle #1$}}}}

\def\e{\mathrm{ e}}

\def\j{\mathrm{ j}}

\def\dB{\,\mathrm{ dB}}

\newcommand{\argmax}{\mathop{\mathrm{argmax}}}

\newcolumntype{C}[1]{>{\centering\arraybackslash}p{#1}}
\def\nr{\mathrm{nr}}

\begin{document}

	
\title{Resampling Images to a Regular Grid from a Non-Regular Subset of Pixel Positions Using Frequency Selective Reconstruction}
\author{J{\"u}rgen~Seiler, \emph{IEEE Member}, Markus Jonscher, Michael Sch{\"o}berl, and~Andr{\'e}~Kaup, \emph{IEEE Fellow}%
\thanks{J.\ Seiler, M.\ Jonscher, and A.\ Kaup are with the Chair of Multimedia Communications and Signal Processing, Friedrich-Alexander-Universit{\"a}t Erlangen-N{\"u}rnberg (FAU), Cauerstr. 7, 91058 Erlangen, Germany (e-mail: juergen.seiler@FAU.de; markus.jonscher@FAU.de; \mbox{andre.kaup@FAU.de}). M.\ Sch{\"o}berl is with the Fraunhofer Institute for Integrated Circuits, Am Wolfsmantel 33, 91058 Erlangen, Germany (e-mail: michael.schoeberl@iis.fraunhofer.de).} \vspace{-0.5cm}}%

\markboth{}%
{Seiler, Jonscher, Sch{\"o}berl, Kaup: Frequency Selective Reconstruction}

\maketitle


\begin{abstract} \label{abstract}
Even though image signals are typically defined on a regular two-dimensional grid, there also exist many scenarios where this is not the case and the amplitude of the image signal only is available for a non-regular subset of pixel positions. In such a case, a resampling of the image to a regular grid has to be carried out. This is necessary since almost all algorithms and technologies for processing, transmitting or displaying image signals rely on the samples being available on a regular grid. Thus, it is of great importance to reconstruct the image on this regular grid so that the reconstruction comes closest to the case that the signal has been originally acquired on the regular grid. In this paper, Frequency Selective Reconstruction is introduced for solving this challenging task. This algorithm reconstructs image signals by exploiting the property that small areas of images can be represented sparsely in the Fourier domain. By further taking into account the basic properties of the Optical Transfer Function of imaging systems, a sparse model of the signal is iteratively generated. In doing so, the proposed algorithm is able to achieve a very high reconstruction quality, in terms of PSNR and SSIM as well as in terms of visual quality. Simulation results show that the proposed algorithm is able to outperform state-of-the-art reconstruction algorithms and gains of more than 1 dB PSNR are possible.

\end{abstract}


\section{Introduction} \label{sec:introduction}
\IEEEPARstart{W}{henever} digital images are considered, it becomes apparent that they are typically defined on a regular two-dimensional grid. That is to say, the pixels are arranged in a rectangular matrix. In many cases, but not necessarily, this regular arrangement directly results from the acquisition process. Aside from this, the positioning of the pixels on a regular grid also is important for displaying images, and especially for processing images.

However, there also exist many scenarios where the amplitudes of an image are not available on a regular rectangular grid, but rather only for a non-regular subset of pixel positions. This might be implicitly caused by the acquisition system as for example in the Optical Cluster Eye \cite{Meyer2011} or the Micro-Optical Artificial Compound Eyes \cite{Duparre2006} which both aim at measuring the light field. Besides this, directly sampling an image at non-regular positions can also be used intentionally in order to reduce the visible influence of aliasing \cite{Hennenfent2007, Maeda2009} or in order to increase the spatial resolution of an imaging sensor \cite{Schoberl2011a} and therewith achieve some kind of super-resolution. For this, it is proposed in \cite{Schoberl2011a} to shield three quarters of every pixel on the sensor non-regularly by an L-shaped mask of varying orientation. This leads to an acquisition of the image on a non-regular grid that has four times the effective resolution as the originally underlying imaging sensor. Aside from this, many other super-resolution algorithms \cite{Park2003} also cause that the final image is not completely available, but rather only a small number of pixels is available that form a non-regular subset of positions with respect to the desired high-resolution grid. In addition to this, there exist various other applications where the image information is not available on a regular two-dimensional grid, but rather only for a non-regular subset of positions.

\begin{figure}
	\centering
	\psfrag{Resampling}[c][c][0.75]{Resampling}
	\psfrag{Non-uniform}[c][c][0.75]{Non-regular subset of pixels}
	\psfrag{Regular}[c][c][0.75]{Image on regular grid}
	\includegraphics[width=0.48\textwidth]{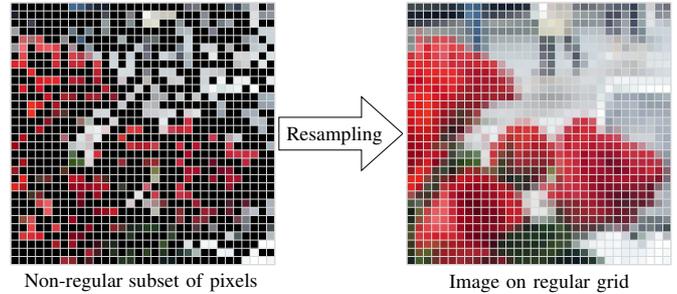}
	\caption{Example for the task of resampling images to a regular grid. Left: Pixel information only available for a non-regular subset of pixel positions. Right: Resampled image defined on a regular grid. }
	\label{fig:sampling_example}
\end{figure}

Independent of the actual reason for the pixels being only available for a non-regular subset of positions, for further processing such signals or displaying them, they have to be resampled to a regular grid as most signal processing algorithms cannot operate on non-regularly spaced data. This task exemplarily is shown in Figure \ref{fig:sampling_example} where on the left side an image is given whose pixels are only available on a non-regular subset of positions. The task of the resampling is then to recover the image on a regular grid, which is shown on the right side. In this article, the Frequency Selective Reconstruction (FSR) algorithm will be introduced as a method for resampling images to a regular grid from a non-regular subset of pixel positions. The algorithm is based on Frequency Selective Extrapolation \cite{Seiler2010c} and exploits the existence of sparse Fourier domain representations for image signals for the reconstruction.

The property of image signals that they can be sparsely represented in different domains is a very fundamental one and has been widely used in very different ways. One application where it has become very popular in the recent years is Compressed Sensing (CS) \cite{Candes2006, Donoho2006} where the sparsity property directly is exploited during the acquisition process. Aside from this, sparsity is an often used property for solving various signal processing tasks \cite{Starck2010,Elad2010a} as for example the denoising of signals \cite{Elad2006} or the deconvolution of signals \cite{Bronstein2005}.

The remainder of the article is organized as follows. In the next section, an overview of various algorithms that can be used for resampling images to a regular grid from a non-regular subset of pixel positions is provided. Afterwards, the proposed reconstruction algorithm is introduced. This is followed by a section discussing how the proposed algorithm can be interpreted within the CS framework, before the performance of the proposed algorithm is demonstrated in \mbox{Section \ref{sec:simulations}}. For assessing the reconstruction quality of the proposed algorithm, this section also gives an extensive comparison to alternative algorithms. Finally, the paper closes with a summary and an outlook to further applications of the reconstruction algorithm.


\section{State-of-the-art Algorithms for Resampling Images to a Regular Grid} \label{sec:nu_sampling} 
Independent of the actual reason for the image information being only available on a non-regular subset of pixel positions, for further processing or displaying, a resampling of the signal to a regular grid is necessary. For this, the scenario is regarded that the available non-regularly spaced signal samples $s_\nr\left[x,y\right]$ result from subsampling at non-regularly spaced positions an unavailable signal $s\left[x,y\right]$ which is defined on a fine discrete regular grid. In this context, $x$ and $y$ depict the spatial variables. In doing so, the available signal $s_\nr\left[x,y\right]$ is defined only on a non-regular subset of the positions with respect to the unavailable signal $s\left[x,y\right]$. Depending on the generation process of the available signal, it may also be possible that the samples are located at spatially continuous positions. In such a case, a quantization of the spatial positions to the fine discrete grid would be necessary. Independent of the actual origin,
\begin{equation}
\label{eq:non_reg_generation}
s_\nr\left[x,y\right] = s\left[x,y\right] b\left[x,y\right]
\end{equation}
can be regarded as generation process of the available signal with non-regularly spaced pixels. The subsampling mask $b\left[x,y\right]$ is one for all available samples and zero for all other samples. The resampling process now aims at reconstructing $s\left[x,y\right]$ in the best possible way, based only on the samples available in $s_\nr\left[x,y\right]$ and the known subsampling mask $b\left[x,y\right]$.

In literature, there exist many different ways for addressing this under-determined signal processing problem. One possibility is to assume that the underlying signal was band-limited to half of the average subsampling frequency, whereas the latter is determined by the inverse of the average distance between the available pixel positions. If this condition was fulfilled, a perfect reconstruction would be possible \cite{Maymon2011, Grochenig1992} in the same way as the reconstruction of regularly subsampled band-limited signals. The algorithms from Papoulis and Gerchberg \cite{Papoulis1975, Gerchberg1974} can be regarded as two very early ones to solve this problem. For this, they iteratively determine a solution that fulfills the band limitation and at the same time fits the sampled values. Besides these algorithms, many more \cite{Early1997, Grochenig1992, Yaroslavsky2009} have been proposed which also use this assumption for the reconstruction but achieve the solution in different ways. The reconstruction in shift-invariant spaces \cite{Eldar2009,Aldroubi2001} can be regarded as a generalization for the reconstruction of band-limited signals but with similar strict requirements on the signal properties. Independent of the actually considered algorithm, it can be discovered that the reconstruction performance drops if the underlying assumption of a signal band-limited to half the average sampling frequency is hurt.

Besides this group of algorithms, several others exist. For achieving a very fast and computational efficient reconstruction, the Four-Nearest-Neighbors Interpolation \cite{Ramponi2001} can be used. This algorithm uses the weighted average of the four nearest neighboring samples as estimate for the unknown signal, while the weights depend on the distance to the known neighbors. In a similar way, the Natural-Neighbor Interpolation \cite{Sibson1981} also performs a weighted averaging of neighboring samples, but with weights for the known samples based on a Voronoi tessellation.  A weighted averaging also is used in the algorithm from \cite{Vazquez2000} as intermediate step for a reconstruction in the Wavelet domain. In addition to these, many classical image reconstruction techniques like patch-based inpainting \cite{Bertalmio2000} can be used for estimating the unknown samples on the fine grid, as well.

An alternative approach is the approximation of the available samples with polynomial functions \cite{LeFloch1996}. In this case, the impact of the different functions decreases with increasing distance to the available samples. In \cite{Wolberg1997}, multilevel \mbox{B-splines} are used for generating a continuous surface which approximates the sampling points. Hence, for recovering the signal on the regular grid, this continuous function would have to be evaluated at the corresponding positions. 

A technique which is often used for resampling an image to a regular grid is variational calculus. Two algorithms which make use of this concept are the ones from \cite{Bughin2008, Vazquez2002}. There, the smoothness of the estimated signal is used as regularization term during the reconstruction process. A reconstruction framework making use of total variation minimization has been proposed in \cite{Dahl2010}. The variational calculus concepts can further be combined with a reconstruction based on spline interpolation \cite{Almansa2010}. A different algorithm which can also be used for handling total variation regularization based problems is the Constrained Split Augmented Lagrangian Shrinkage Algorithm \cite{Afonso2011}.
	 
Aside from these, statistical modeling can also be used for estimating the samples on the regular grid. In \cite{Takeda2007}, a statistical modeling framework is proposed which can use non-parametric kernel regression or a steering kernel regression for recovering the signal on a regular grid. An alternative method is presented in \cite{Zhai2012} where a hybrid image reconstruction algorithm is proposed that uses parametric and non-parametric modeling of the image signals. There, the strengths of both approaches are put together in a multi-scale framework for image reconstruction.

As mentioned in the previous section, the property that sparse representations exist for image signals can be exploited in very different areas and hence also for the resampling of images to a regular grid. Two of the algorithms that exploit this property are the sparse reconstruction with operator learning from \cite{Hawe2013} and the Morphological Component Analysis from \cite{Elad2005}. There, the objective is to decompose the image signal into its underlying atomic functions based only on the available samples.


\section{Resampling Images to a Regular Grid by Frequency Selective Reconstruction} \label{sec:fse_reconstruction}
As has been shown in the previous section, in literature, there already exist many algorithms for recovering image signals on a regular grid from a non-regular subset of pixel positions. The objective of any algorithm for reconstructing an image signal on a fine regular grid can be regarded as follows. Considering the generation model (\ref{eq:non_reg_generation}), the available signal $s_\nr\left[x,y\right]$ results from subsampling the desired signal $s\left[x,y\right]$ on a non-regular grid while the desired signal is defined on a fine regular grid. Thus, the resampling process always aims at generating $\hat{s}\left[x,y\right]$ as estimate for the unknown signal $s\left[x,y\right]$ in such a way that the error becomes as small as possible.

The algorithms mentioned above have quite different origins and exploit different signal properties for solving the under-determined problem of estimating the missing samples on the fine regular grid. However, unfortunately most of the algorithms mentioned above are only able to recover low frequency content at a high quality and have problems if the signal to be recovered contains high frequency content. In this article, Frequency Selective Reconstruction (FSR) is introduced which exploits the property that for image signals sparse representations in the Fourier domain exist. In doing so, FSR is able to even resample images with high frequency content. Before FSR is outlined in the second half of this section in detail, the impact of non-regular subsampling on the spectrum of a signal and the general reconstruction principle of FSR is discussed in the next subsection.

\subsection{Consequences from Non-Regular Subsampling and Frequency Selective Reconstruction Principle} \label{ssec:reconstruction_principle}

If image signals are regarded, it is widely known that they can be sparsely represented in the Fourier domain \cite{Candes2007}. That is to say that most signal energy is concentrated into a small number of transform coefficients while all other coefficients are equal or close to zero. However, since the content of images can vary strongly, image signals are quite non-stationary and the sparsity assumption only holds if image patches are considered where the signal can be regarded as stationary. In order to account for this, the later introduced FSR performs a block-wise processing of the image as will be shown in the next subsection.

Hence, for discussing the reconstruction principle, the image block $f\left[m,n\right]$ of size $M\times N$ and with spatial coordinates $m$ and $n$ is considered. This block is located at position $\left(x_o, y_o\right)$ in the image signal $s\left[x,y\right]$ which is defined on a regular grid. The block can be extracted from the image signal according to
\begin{equation}
f\left[m,n\right] = s\left[x_o+m, y_o+n\right] .
\end{equation}
Signal $f\left[m,n\right]$ can also be decomposed into Fourier basis functions and can be written as
\begin{equation}
f\left[m,n\right] = \sum_{\left(k,l\right)\in\mathcal{D}} c_{\left(k,l\right)} \varphi_{\left(k,l\right)}\left[m,n\right]
\end{equation}
with the Fourier basis functions
\begin{equation}
\label{eq:fourier_bf}
\varphi_{\left(k,l\right)}\left[m,n\right] = \e^{\frac{2\pi\j}{M}km} \e^{\frac{2\pi\j}{N}ln}
\end{equation}
and the corresponding expansion coefficients $c_{\left(k,l\right)}$. Set $\mathcal{D}$ subsumes the indices $k,l$ of all possible basis functions. If $f\left[m,n\right]$ is sparse in the Fourier domain, most $c_{\left(k,l\right)}$ are equal to zero or contribute only negligibly to the signal. 

If the corresponding block $f_\nr\left[m,n\right]$ from the non-regularly subsampled image $s_\nr\left[x,y\right]$ is regarded, its generation process can be described similarly to (\ref{eq:non_reg_generation}) by 
\begin{equation}
\label{eq:non_reg_generation_block}
f_\nr\left[m,n\right] =  f\left[m,n\right] q\left[m,n\right]
\end{equation}
with $q\left[m,n\right]$ being the corresponding block from mask $b\left[x,y\right]$. Taking a look at the Fourier-transform of $f\left[m,n\right]$, the corresponding spectrum 
\begin{equation}
F\left[k,l\right] =  \mathcal{F}\left\{f\left[m,n\right]\right\} = \sum_{\forall\left(m,n\right)} f\left[m,n\right] \varphi^\ast_{\left(k,l\right)}\left[m,n\right]
\end{equation}
results from the scalar product between the spatial domain signal and the basis functions. If the generation process (\ref{eq:non_reg_generation_block}) of the non-regular subsampling is taken into account, the spectrum 
\begin{eqnarray}
F_\nr\left[k,l\right] &=&  \mathcal{F}\left\{f_\nr\left[m,n\right]\right\} \\
&=& \frac{1}{MN} F\left[k,l\right] \circledast Q\left[k,l\right]
\end{eqnarray}
of the non-regularly subsampled signal can be defined accordingly, but can also be regarded as the two-dimensional circular convolution (denoted by $\circledast$) of spectrum $F\left[k,l\right]$ with the Fourier-transform 
\begin{equation}
Q\left[k,l\right] = \mathcal{F}\left\{q\left[m,n\right]\right\}
\end{equation}
of the mask  $q\left[m,n\right]$. 

\begin{figure}
\psfrag{qmn}[l][l][0.75]{$q\left[m,n\right]$}
\psfrag{Qkl}[l][l][0.75]{$\left|Q\left[k,l\right]\right|$}
\psfrag{k}[l][l][0.75]{$k$}
\psfrag{l}[r][l][0.75]{$l$}
\psfrag{m}[l][l][0.75]{$m$}
\psfrag{n}[l][l][0.75]{$n$}
\psfrag{x01}[t][t][0.5]{$0$}%
\psfrag{x02}[t][t][0.5]{$5$}%
\psfrag{x03}[t][t][0.5]{$10$}%
\psfrag{x04}[t][t][0.5]{$15$}%
\psfrag{x05}[t][t][0.5]{$20$}%
\psfrag{x06}[t][t][0.5]{$25$}%
\psfrag{x07}[t][t][0.5]{$30$}%
\psfrag{v01}[r][r][0.5]{$0$}%
\psfrag{v02}[r][r][0.5]{$10$}%
\psfrag{v03}[r][r][0.5]{$20$}%
\psfrag{v04}[r][r][0.5]{$30$}%
\psfrag{z01}[r][r][0.5]{}%
\psfrag{z02}[r][r][0.5]{$0.2$}%
\psfrag{z03}[r][r][0.5]{}%
\psfrag{z04}[r][r][0.5]{$0.6$}%
\psfrag{z05}[r][r][0.5]{}%
\psfrag{z06}[r][r][0.5]{$1$}%
\includegraphics[width=0.46\textwidth]{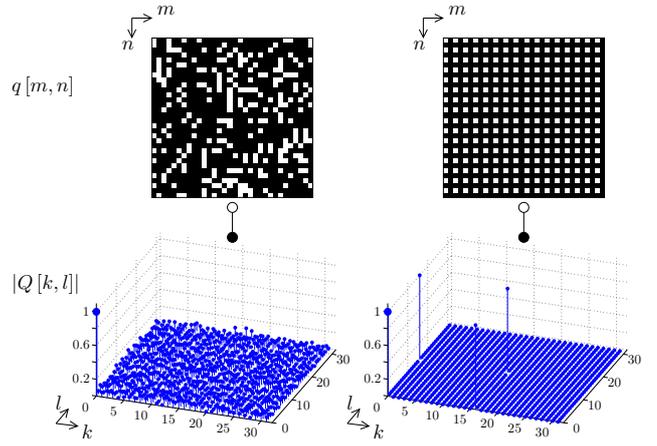}
\caption{Comparison of a non-regular subsampling mask with a density of $25\%$ and its corresponding absolute spectrum (normalized to maximum) to a regular subsampling with density $25\%$. }
\label{fig:sampling_spectra_comparison}
\end{figure}

Regarding the spectrum of the mask $q\left[m,n\right]$, it can be seen that the spectrum exhibits a dominant peak at frequency zero while the contribution of all other frequencies is very small and can be regarded as a noise-like floor. In Figure \ref{fig:sampling_spectra_comparison}, a mask for a non-regular subsampling with a density of $25\%$ is shown together with its corresponding spectrum. There, the dominant peak at the origin as well as the noise-like floor can be seen quite well. The level of the noise-like floor would increase, or respectively decrease, according to the subsampling density, while the main peak always remains. In addition to the non-regular subsampling pattern, the figure also shows a regular subsampling pattern with a density of $25\%$ which would be equal to a reduction of the number of samples by a factor of $2$ in horizontal and vertical direction. This spectrum exhibits four equally strong peaks at zero and half the maximum frequencies in both directions. These four peaks would cause the aliasing in the case of a regular subsampling of the signal $f\left[m,n\right]$, resulting in the fact that if the original spectrum $F\left[m,n\right]$ was convolved with the spectrum of this mask, ambiguities occur and the different frequencies get superimposed on each other. This especially harms high frequencies which are superimposed by the typically more dominant low frequencies, heavily distorting fine structures. 

\begin{figure}
\psfrag{qmn}[l][l][0.75]{$q\left[m,n\right]$}
\psfrag{Qkl}[l][l][0.75]{$\left|Q\left[k,l\right]\right|$}
\psfrag{fmn}[l][l][0.75]{$f\left[m,n\right]$}
\psfrag{Fkl}[l][l][0.75]{$\left|F\left[k,l\right]\right|$}
\psfrag{fnrmn}[l][l][0.75]{$f_\nr\left[m,n\right]$}
\psfrag{Fnrkl}[l][l][0.75]{$\left|F_\nr\left[k,l\right]\right|$}
\psfrag{k}[l][l][0.75]{$k$}
\psfrag{l}[r][l][0.75]{$l$}
\psfrag{m}[l][l][0.75]{$m$}
\psfrag{n}[l][l][0.75]{$n$}
\psfrag{astast}[l][l][1.]{$\circledast$}

\psfrag{x01}[t][t][0.5]{$0$}%
\psfrag{x02}[t][t][0.5]{$5$}%
\psfrag{x03}[t][t][0.5]{$10$}%
\psfrag{x04}[t][t][0.5]{$15$}%
\psfrag{x05}[t][t][0.5]{$20$}%
\psfrag{x06}[t][t][0.5]{$25$}%
\psfrag{x07}[t][t][0.5]{$30$}%
\psfrag{v01}[r][r][0.5]{$0$}%
\psfrag{v02}[r][r][0.5]{$10$}%
\psfrag{v03}[r][r][0.5]{$20$}%
\psfrag{v04}[r][r][0.5]{$30$}%
\psfrag{z01}[r][r][0.5]{}%
\psfrag{z02}[r][r][0.5]{$0.2$}%
\psfrag{z03}[r][r][0.5]{}%
\psfrag{z04}[r][r][0.5]{$0.6$}%
\psfrag{z05}[r][r][0.5]{}%
\psfrag{z06}[r][r][0.5]{$1$}%

\includegraphics[width=0.46\textwidth]{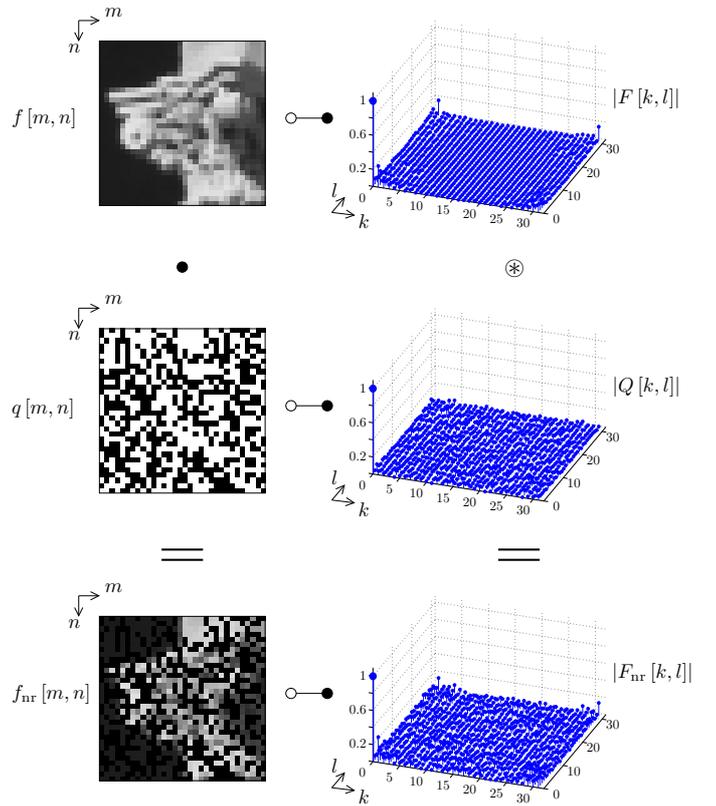}
\caption{Impact of non-regular subsampling on image patch $f\left[m,n\right]$ and the corresponding spectrum $F\left[k,l\right]$. }
\label{fig:nr_sampling_principle}
\end{figure}

However, in case of a non-regular subsampling, a different aliasing occurs than in the case of regular subsampling. While regular subsampling causes a repetition of the original spectrum shifted according to the subsampling frequency, in the case of non-regular subsampling the aliasing looks similar to a noise-like contribution in the frequency domain. That is to say, every originally present frequency causes a small additional contribution to all other frequencies. This property becomes very important if considered together with the assumption that the image patch can be sparsely represented in the Fourier domain. If this is true, the dominant basis functions survive the subsampling process and can still be identified in the available signal $f_\nr\left[m,n\right]$ of non-regularly spaced samples. To illustrate this, Figure \ref{fig:nr_sampling_principle} shows the impact of non-regular subsampling on the signal $f\left[m,n\right]$ and the corresponding spectrum $F\left[k,l\right]$. As the generation process for a non-regular subsampling consists of the multiplication of the theoretical signal on the fine grid with the subsampling mask, the original spectrum gets convolved with the spectrum of the subsampling mask. However, as long as the original signal is sparse in the Fourier domain, the dominant frequencies, or respectively, basis functions remain dominant in the non-regularly subsampled signal. This can also be discovered in the figure. Even though the non-regular subsampling causes a strong noise-like floor in the spectrum $F\left[k,l\right]$, the frequencies which were strongly present in the original spectrum still tower above this floor.

Thus, the reconstruction process has to aim at identifying the dominant frequencies in the spectrum of the non-regularly subsampled signal and estimate the appropriate weights. This is a challenging task, since the amplitude of every frequency in the transformed non-regularly subsampled signal consists of two portions. This is always the actual weight of the regarded frequency and the contributions from all other frequencies, superimposed due to the convolution of the original spectrum with the transformed mask. In the case that the spectrum exhibits very strong frequencies, it may occur that their impact on other frequencies becomes larger than the actual weights of the latter. In order to account for this, we propose to use an iterative procedure for the reconstruction in which in every iteration the strongest present frequency is determined and its weight is estimated. Afterwards, the influence of this frequency on all other frequencies can be taken into account and therewith reducing the superimposed aliasing-like noise term for the estimation of the successive components.

In this context, it has to be noted that the objective of the reconstruction is not to model the available non-regularly spaced samples best by generating the sparse model. Instead, the reconstruction aims at estimating the original frequency weights as good as possible. A way how to exploit this effectively is outlined in the next subsection where the proposed algorithm is presented in detail.

\subsection{Frequency Selective Reconstruction Algorithm} \label{ssec:reconstruction_algorithm}

The now introduced Frequency Selective Reconstruction (FSR) is based on Frequency Selective Extrapolation \cite{Meisinger2004b} which is a quite general approach for signal extrapolation. This algorithm has received several enhancements \cite{Seiler2008, Seiler2010c, Seiler2011c} in order to improve the extrapolation quality and has been applied to several signal extrapolation tasks in the area of image and video signal processing. To name just a few, this has been applications like error concealment in video communication \cite{Seiler2011} or defect pixel compensation \cite{Schoberl2011}. However, unfortunately Frequency Selective Extrapolation is not suited that well for the considered resampling task. Thus, the proposed FSR incorporates a novel processing order, which accounts for the local density of the available samples and, even more important, considers the Optical Transfer Function of imaging systems for avoiding ambiguities during the modeling.

For resampling an image from a non-regular subset of pixel positions to a regular grid, FSR splits the image into blocks which are processed consecutively. In doing so, the reconstruction can account for the instationarity of typical image signals and the property that an image block can be sparsely represented in the Fourier-domain can be exploited. Additionally, by processing the image blocks one after the other, the content of one image block can be utilized for the reconstruction of the next block and therewith improving the overall reconstruction quality.

\begin{figure}
	\centering
	\psfrag{m}[l][l][1]{$x$}
	\psfrag{n}[r][l][1]{$y$}
	\psfrag{bmn}[c][c][1]{$b\left[x,y\right]$}%
	\psfrag{btmn}[c][c][1]{$\widetilde{b}\left[x,y\right]$}%
	\psfrag{procorder}[c][c][0.8]{Processing order}%
	
\psfrag{s05}[c][c][0.4]{$1$}%
\psfrag{s06}[c][c][0.4]{$2$}%
\psfrag{s07}[c][c][0.4]{$3$}%
\psfrag{s08}[c][c][0.4]{$4$}%
\psfrag{s09}[c][c][0.4]{$5$}%
\psfrag{s10}[c][c][0.4]{$6$}%
\psfrag{s11}[c][c][0.4]{$7$}%
\psfrag{s12}[c][c][0.4]{$8$}%
\psfrag{s13}[c][c][0.4]{$9$}%
\psfrag{s14}[c][c][0.4]{$10$}%
\psfrag{s15}[c][c][0.4]{$11$}%
\psfrag{s16}[c][c][0.4]{$12$}%
\psfrag{s17}[c][c][0.4]{$13$}%
\psfrag{s18}[c][c][0.4]{$14$}%
\psfrag{s19}[c][c][0.4]{$15$}%
\psfrag{s20}[c][c][0.4]{$16$}%
\psfrag{s21}[c][c][0.4]{$17$}%
\psfrag{s22}[c][c][0.4]{$18$}%
\psfrag{s23}[c][c][0.4]{$19$}%
\psfrag{s24}[c][c][0.4]{$20$}%
\psfrag{s25}[c][c][0.4]{$21$}%
\psfrag{s26}[c][c][0.4]{$22$}%
\psfrag{s27}[c][c][0.4]{$23$}%
\psfrag{s28}[c][c][0.4]{$24$}%
\psfrag{s29}[c][c][0.4]{$25$}%
\psfrag{s30}[c][c][0.4]{$26$}%
\psfrag{s31}[c][c][0.4]{$27$}%
\psfrag{s32}[c][c][0.4]{$28$}%
\psfrag{s33}[c][c][0.4]{$29$}%
\psfrag{s34}[c][c][0.4]{$30$}%
\psfrag{s35}[c][c][0.4]{$31$}%
\psfrag{s36}[c][c][0.4]{$32$}%
\psfrag{s37}[c][c][0.4]{$33$}%
\psfrag{s38}[c][c][0.4]{$34$}%
\psfrag{s39}[c][c][0.4]{$35$}%
\psfrag{s40}[c][c][0.4]{$36$}%
\psfrag{s41}[c][c][0.4]{$37$}%
\psfrag{s42}[c][c][0.4]{$38$}%
\psfrag{s43}[c][c][0.4]{$39$}%
\psfrag{s44}[c][c][0.4]{$40$}%
\psfrag{s45}[c][c][0.4]{$41$}%
\psfrag{s46}[c][c][0.4]{$42$}%
\psfrag{s47}[c][c][0.4]{$43$}%
\psfrag{s48}[c][c][0.4]{$44$}%
\psfrag{s49}[c][c][0.4]{$45$}%
\psfrag{s50}[c][c][0.4]{$46$}%
\psfrag{s51}[c][c][0.4]{$47$}%
\psfrag{s52}[c][c][0.4]{$48$}%
\psfrag{s53}[c][c][0.4]{$49$}%
\psfrag{s54}[c][c][0.4]{$50$}%
\psfrag{s55}[c][c][0.4]{$51$}%
\psfrag{s56}[c][c][0.4]{$52$}%
\psfrag{s57}[c][c][0.4]{$53$}%
\psfrag{s58}[c][c][0.4]{$54$}%
\psfrag{s59}[c][c][0.4]{$55$}%
\psfrag{s60}[c][c][0.4]{$56$}%
\psfrag{s61}[c][c][0.4]{$57$}%
\psfrag{s62}[c][c][0.4]{$58$}%
\psfrag{s63}[c][c][0.4]{$59$}%
\psfrag{s64}[c][c][0.4]{$60$}%
\psfrag{s65}[c][c][0.4]{$61$}%
\psfrag{s66}[c][c][0.4]{$62$}%
\psfrag{s67}[c][c][0.4]{$63$}%
\psfrag{s68}[c][c][0.4]{$64$}%

	\includegraphics[width=0.45\textwidth]{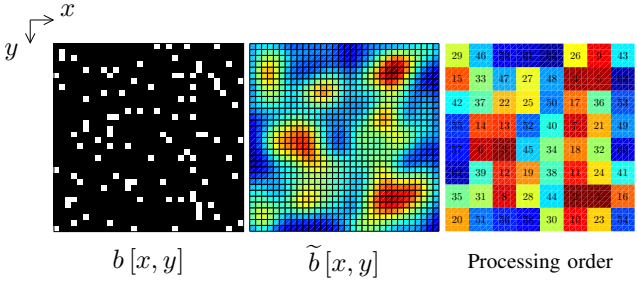}
	\caption{Example for determining processing order. Left: subsampling mask $b\left[x,y\right]$. Mid: filtered mask $\widetilde{b}\left[x,y\right]$. Right: Processing order of image blocks.}
	\label{fig:processing_order}
\end{figure}

In general, there exist many different possibilities for the actual order for processing the image blocks. The most obvious order would be to process the blocks of an image in line scan order. However, this order is not suited well for reusing already extrapolated areas and does not take into account the structure of the signal to be reconstructed. In \cite{Seiler2011c} an optimized processing order is proposed which is better suited for reusing already reconstructed samples and it exhibits a superior behavior over line scan order. However, this processing order was originally designed for extrapolating large areas of missing samples and is not suited that well for the considered scenario of resampling images from non-regularly spaced positions. Thus, a new processing order is proposed which takes the characteristics of the signal reconstruction problem into account. As typically blocks with many known samples can be reconstructed more accurately than blocks with few, the former one should be processed first and therewith support the reconstruction of the latter one. 

In order to account for this, the local density of available samples is considered. Therefore, the subsampling mask $b\left[x,y\right]$ is lowpass filtered with a two-dimensional gaussian window $d\left[x,y\right]$ leading to the filtered mask
\begin{equation}
\widetilde{b}\left[x,y\right] = b\left[x,y\right] \ast d\left[x,y\right] .
\end{equation}
In this process, the half width of the gaussian window is selected to be of the same size as the used block size. Afterwards, all the values of $\widetilde{b}\left[x,y\right]$ within each block are summed up and the blocks get processed in decreasing order of the result of the summation. In order to illustrate the computation of the processing order, Figure \ref{fig:processing_order} shows a small example of a subsampling mask, the corresponding filtered mask and the final processing order. It can be recognized well that areas with many available samples exhibit a high local density, resulting in an early processing whereas blocks with few samples are processed at a later stage. The filtering process is necessary, as for very low subsampling densities it might happen that there exist several neighboring blocks that all contain no known samples. By applying the filtering operation first, it can be assured that these large areas without known samples get closed from the outer margin to the center.

\begin{figure}
	\centering
	\psfrag{m}[l][l][1]{$m$}
	\psfrag{n}[l][l][1]{$n$}
	\psfrag{A}[l][l][1]{$\mathcal{A}$}
	\psfrag{B}[l][l][1]{$\mathcal{B}$}
	\psfrag{R}[l][l][1]{$\mathcal{R}$}
	\psfrag{L}[l][l][1]{$\mathcal{L}=\mathcal{A}\cup\mathcal{B}\cup\mathcal{R}$}
	\psfrag{Border}[c][c][1]{Border}
	\psfrag{width}[c][c][1]{width}
	\psfrag{Block}[c][c][1]{Block}
	\psfrag{size}[c][c][1]{size}
	\includegraphics[width=0.3\textwidth]{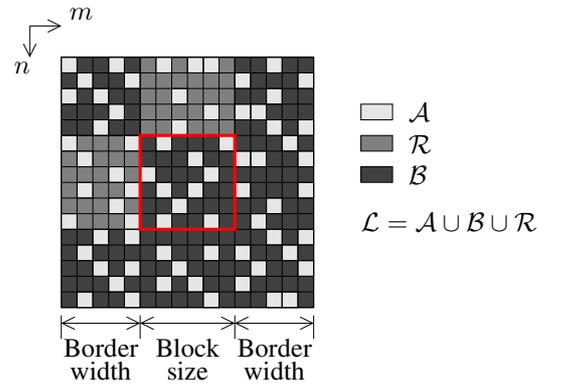}
	\caption{Reconstruction area $\mathcal{L}$ consisting of available samples subsumed in area $\mathcal{A}$, already reconstructed samples subsumed in area $\mathcal{R}$, and unknown samples subsumed in area $\mathcal{B}$. The block actually to be reconstructed is marked in red in the center.}
	\label{fig:extrapolation_area}
\end{figure}

After the processing order of the blocks has been determined, the reconstruction of the individual blocks starts. If an image block is to be reconstructed by FSR, the block actually to be reconstructed always is regarded together with a spatial neighborhood as shown in Figure \ref{fig:extrapolation_area} as an example. This neighborhood is a stripe of samples and the width of this stripe is called border width. Together with its neighborhood, the block forms the so-called reconstruction area $\mathcal{L}$. If the available signal is not of rectangular shape, or respectively, if the reconstruction should be carried out on a larger area, the signal can be padded with samples of arbitrary amplitude, accordingly. All the samples in area $\mathcal{L}$ can be divided into three groups. First, all originally known samples are subsumed in area $\mathcal{A}$ while all unknown samples are subsumed in area $\mathcal{B}$. If values have been padded to extend the signal, these samples also belong to set $\mathcal{B}$. Since all the samples that may have been used for padding the signal are regarded as unknown and do not contribute to the modeling, any amplitude can be assigned to them. As neighboring blocks might already have been processed before, there exists a reconstruction for the unknown samples. These previously reconstructed samples can be used for the reconstruction of the samples in the currently regarded block and are subsumed in area $\mathcal{R}$ of already reconstructed samples. Due to the reuse of already reconstructed samples, the signal in area $\mathcal{L}$ slightly differs from the definition in the preceding subsection and thus is depicted by $\bar{f}_\nr\left[m,n\right]$. Nevertheless, the properties and concepts for the reconstruction do not change.

The basic idea of FSR is to generate the sparse model 
\begin{equation}
 g\left[m,n\right] = \sum_{\left(k,l\right)\in\mathcal{K}}\hat{c}_{\left(k,l \right)} \varphi_{\left(k,l \right)}\left[m,n\right]
\end{equation}
of the signal as weighted superposition of Fourier basis functions, based on the known samples. The basis functions $\varphi_{\left(k,l \right)}\left[m,n\right]$ are defined according to (\ref{eq:fourier_bf}) and $\hat{c}_{\left(k,l \right)}$ depict the corresponding expansion coefficients to be determined. The set $\mathcal{K}$ contains the indices of all basis functions to be used. 

As proposed in the preceding subsection, an iterative procedure should be used for the reconstruction. Accordingly, the model is generated iteratively and in every iteration, one basis function is selected and the corresponding weight is estimated. Initially, the model $g^{\left(0\right)}\left[m,n\right]$ is set to zero. Thus, the residual 
\begin{equation}
r^{\left(\nu\right)} \left[m,n\right] = \bar{f}_\nr\left[m,n\right] - g^{\left(\nu\right)} \left[m,n\right]
\end{equation}
between model and available samples for iteration $\nu$, becomes
\begin{equation}
r^{\left(0\right)} \left[m,n\right] = \bar{f}_\nr\left[m,n\right]
\end{equation}
prior to the first iteration. For generating the model over the iterations, the weighted residual energy
\begin{equation}
E_w =\sum_{\left(m,n\right)\in\mathcal{L}} \left|  \bar{f}_\nr\left[m,n\right] - g\left[m,n\right] \right|^2 w\left[m,n\right]
\end{equation}
is considered. In this context, the weighting function 
\begin{equation}
\label{eq:weighting_function}
w\left[m,n\right]\hspace{-1mm} = \hspace{-1mm}\left\{ \hspace{-1mm}\begin{array}{ll} \hat{\rho}^{\sqrt{\left(m-\frac{M-1}{2}\right)^2+\left(n-\frac{N-1}{2}\right)^2}} & \mbox{for } \left(m,n\right)\in \mathcal{A} \\ \delta \hat{\rho}^{\sqrt{\left(m-\frac{M-1}{2}\right)^2+\left(n-\frac{N-1}{2}\right)^2}} & \mbox{for } \left(m,n\right)\in \mathcal{R} \\ 0 & \mbox{for } \left(m,n\right)\in \mathcal{B}\end{array}\right. ,
\end{equation}
originally introduced in \cite{Meisinger2004b}, is used to assign different weights to different regions. In doing so, it can be achieved that different samples obtain different influence on the model generation, depending on their position. Thus, starting from the center an exponentially decreasing weighting is assigned to all available samples. The speed of the decay is controlled by factor $\hat{\rho}$. As the already reconstructed samples in area $\mathcal{R}$ are not as reliable as originally available ones, their weight is further reduced by a factor $\delta$ from the range between zero and one. Since the samples in $\mathcal{B}$ are unknown, they cannot contribute to the model generation and are weighted by zero, accordingly.

For determining which basis function to add in iteration $\nu$, a weighted projection of the residual $r^{\left(\nu-1\right)}\left[m,n\right]$ from the preceding iteration is carried out onto all basis functions, as proposed in \cite{Meisinger2004b, Seiler2011c}. For this, the weighted residual energy 
\begin{equation}
\label{eq:weighted_residual_iteration}\widetilde{E}_{w,\left(k,l \right)}^{\left(\nu\right)}\hspace{-1mm} =\hspace{-4mm}\sum_{\left(m,n\right)\in\mathcal{L}}\hspace{-3mm} \left|r^{\left(\nu-1\right)}\left[m,n\right] - p_{\left(k,l \right)}^{\left(\nu\right)} \varphi_{\left(k,l \right)}\left[m,n\right] \right|^2 w\left[m,n\right]
\end{equation}
with respect to basis function $\varphi_{\left(k,l \right)}\left[m,n\right]$ is regarded. The projection coefficient $p_{\left(k,l \right)}^{\left(\nu\right)}$ minimizes $\widetilde{E}_{w,\left(k,l \right)}^{\left(\nu\right)}$ for the considered basis function and can be calculated by setting the partial derivatives 
\begin{equation}
\frac{\partial \widetilde{E}_{w,\left(k,l \right)}^{\left(\nu\right)}}{\partial p_{\left(k,l \right)}^{\left(\nu\right)}}\stackrel{!}{=} 0 \hspace{1cm}\mbox{and}\hspace{1cm} \frac{\partial \widetilde{E}_{w,\left(k,l \right)}^{\left(\nu\right)}}{\partial p_{\left(k,l \right)}^{\left(\nu\right)\ast}}\stackrel{!}{=} 0
\end{equation}
to zero. This yields 
\begin{equation}
\label{eq:projection_coefficient}
 p_{\left(k,l \right)}^{\left(\nu\right)} = \frac{\displaystyle \sum_{\left(m,n\right)\in\mathcal{L}} r^{\left(\nu-1\right)} \left[m,n\right] \varphi^\ast_{\left(k,l \right)}\left[m,n\right] w\left[m,n\right]}{\displaystyle \sum_{\left(m,n\right)\in\mathcal{L}} \varphi_{\left(k,l \right)}^\ast\left[m,n\right]w\left[m,n\right]\varphi_{\left(k,l \right)}\left[m,n\right]} 
\end{equation}
which is calculated for all $\left(k,l\right)$. 

As shown in \cite{Seiler2010c}, the calculations can be efficiently implemented in the frequency domain. Thus (\ref{eq:projection_coefficient}) can be expressed in the frequency domain by
\begin{equation}
 p_{\left(k,l \right)}^{\left(\nu\right)} = \frac{R_w^{\left(\nu-1\right)}\left[k,l\right]}{W\left[0,0\right]}
\end{equation}
with $R_w^{\left(\nu-1\right)}\left[k,l\right]$ being the Fourier-transform of the weighted residual
\begin{equation}
 r_w^{\left(\nu-1\right)}\left[m,n\right] =  r^{\left(\nu-1\right)}\left[m,n\right]\cdot w\left[m,n\right]
\end{equation}
and $W\left[k,l\right]$ the transformed weighting function.

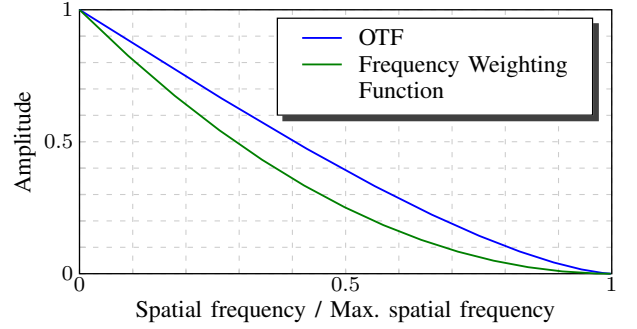
\begin{figure}
	\centering



\providelength{\AxesLineWidth}       \setlength{\AxesLineWidth}{0.5pt}
\providelength{\GridLineWidth}       \setlength{\GridLineWidth}{0.4pt}
\providelength{\GridLineDotSep}      \setlength{\GridLineDotSep}{0.4pt}
\providelength{\MinorGridLineWidth}  \setlength{\MinorGridLineWidth}{0.4pt}
\providelength{\MinorGridLineDotSep} \setlength{\MinorGridLineDotSep}{0.8pt}
\providelength{\plotwidth}           \setlength{\plotwidth}{7cm} 
\providelength{\LineWidth}           \setlength{\LineWidth}{0.7pt}
\providelength{\MarkerSize}          \setlength{\MarkerSize}{4pt}
\newrgbcolor{GridColor}{0.8 0.8 0.8}

\psset{xunit=1.000000\plotwidth,yunit=0.500000\plotwidth}
\begin{pspicture}(-0.107143,-0.133929)(1.000000,1.000000)


\psline[linestyle=dashed,dash=2pt 3pt,dotsep=\GridLineDotSep,linewidth=\GridLineWidth,linecolor=GridColor](0.000000,0.000000)(0.000000,1.000000)
\psline[linestyle=dashed,dash=2pt 3pt,dotsep=\GridLineDotSep,linewidth=\GridLineWidth,linecolor=GridColor](0.100000,0.000000)(0.100000,1.000000)
\psline[linestyle=dashed,dash=2pt 3pt,dotsep=\GridLineDotSep,linewidth=\GridLineWidth,linecolor=GridColor](0.200000,0.000000)(0.200000,1.000000)
\psline[linestyle=dashed,dash=2pt 3pt,dotsep=\GridLineDotSep,linewidth=\GridLineWidth,linecolor=GridColor](0.300000,0.000000)(0.300000,1.000000)
\psline[linestyle=dashed,dash=2pt 3pt,dotsep=\GridLineDotSep,linewidth=\GridLineWidth,linecolor=GridColor](0.400000,0.000000)(0.400000,1.000000)
\psline[linestyle=dashed,dash=2pt 3pt,dotsep=\GridLineDotSep,linewidth=\GridLineWidth,linecolor=GridColor](0.500000,0.000000)(0.500000,1.000000)
\psline[linestyle=dashed,dash=2pt 3pt,dotsep=\GridLineDotSep,linewidth=\GridLineWidth,linecolor=GridColor](0.600000,0.000000)(0.600000,1.000000)
\psline[linestyle=dashed,dash=2pt 3pt,dotsep=\GridLineDotSep,linewidth=\GridLineWidth,linecolor=GridColor](0.700000,0.000000)(0.700000,1.000000)
\psline[linestyle=dashed,dash=2pt 3pt,dotsep=\GridLineDotSep,linewidth=\GridLineWidth,linecolor=GridColor](0.800000,0.000000)(0.800000,1.000000)
\psline[linestyle=dashed,dash=2pt 3pt,dotsep=\GridLineDotSep,linewidth=\GridLineWidth,linecolor=GridColor](0.900000,0.000000)(0.900000,1.000000)
\psline[linestyle=dashed,dash=2pt 3pt,dotsep=\GridLineDotSep,linewidth=\GridLineWidth,linecolor=GridColor](1.000000,0.000000)(1.000000,1.000000)
\psline[linestyle=dashed,dash=2pt 3pt,dotsep=\GridLineDotSep,linewidth=\GridLineWidth,linecolor=GridColor](0.000000,0.000000)(1.000000,0.000000)
\psline[linestyle=dashed,dash=2pt 3pt,dotsep=\GridLineDotSep,linewidth=\GridLineWidth,linecolor=GridColor](0.000000,0.100000)(1.000000,0.100000)
\psline[linestyle=dashed,dash=2pt 3pt,dotsep=\GridLineDotSep,linewidth=\GridLineWidth,linecolor=GridColor](0.000000,0.200000)(1.000000,0.200000)
\psline[linestyle=dashed,dash=2pt 3pt,dotsep=\GridLineDotSep,linewidth=\GridLineWidth,linecolor=GridColor](0.000000,0.300000)(1.000000,0.300000)
\psline[linestyle=dashed,dash=2pt 3pt,dotsep=\GridLineDotSep,linewidth=\GridLineWidth,linecolor=GridColor](0.000000,0.400000)(1.000000,0.400000)
\psline[linestyle=dashed,dash=2pt 3pt,dotsep=\GridLineDotSep,linewidth=\GridLineWidth,linecolor=GridColor](0.000000,0.500000)(1.000000,0.500000)
\psline[linestyle=dashed,dash=2pt 3pt,dotsep=\GridLineDotSep,linewidth=\GridLineWidth,linecolor=GridColor](0.000000,0.600000)(1.000000,0.600000)
\psline[linestyle=dashed,dash=2pt 3pt,dotsep=\GridLineDotSep,linewidth=\GridLineWidth,linecolor=GridColor](0.000000,0.700000)(1.000000,0.700000)
\psline[linestyle=dashed,dash=2pt 3pt,dotsep=\GridLineDotSep,linewidth=\GridLineWidth,linecolor=GridColor](0.000000,0.800000)(1.000000,0.800000)
\psline[linestyle=dashed,dash=2pt 3pt,dotsep=\GridLineDotSep,linewidth=\GridLineWidth,linecolor=GridColor](0.000000,0.900000)(1.000000,0.900000)
\psline[linestyle=dashed,dash=2pt 3pt,dotsep=\GridLineDotSep,linewidth=\GridLineWidth,linecolor=GridColor](0.000000,1.000000)(1.000000,1.000000)

\psline[linewidth=\AxesLineWidth,linecolor=GridColor](0.000000,0.000000)(0.000000,0.015000)
\psline[linewidth=\AxesLineWidth,linecolor=GridColor](0.100000,0.000000)(0.100000,0.015000)
\psline[linewidth=\AxesLineWidth,linecolor=GridColor](0.200000,0.000000)(0.200000,0.015000)
\psline[linewidth=\AxesLineWidth,linecolor=GridColor](0.300000,0.000000)(0.300000,0.015000)
\psline[linewidth=\AxesLineWidth,linecolor=GridColor](0.400000,0.000000)(0.400000,0.015000)
\psline[linewidth=\AxesLineWidth,linecolor=GridColor](0.500000,0.000000)(0.500000,0.015000)
\psline[linewidth=\AxesLineWidth,linecolor=GridColor](0.600000,0.000000)(0.600000,0.015000)
\psline[linewidth=\AxesLineWidth,linecolor=GridColor](0.700000,0.000000)(0.700000,0.015000)
\psline[linewidth=\AxesLineWidth,linecolor=GridColor](0.800000,0.000000)(0.800000,0.015000)
\psline[linewidth=\AxesLineWidth,linecolor=GridColor](0.900000,0.000000)(0.900000,0.015000)
\psline[linewidth=\AxesLineWidth,linecolor=GridColor](1.000000,0.000000)(1.000000,0.015000)
\psline[linewidth=\AxesLineWidth,linecolor=GridColor](0.000000,0.000000)(0.012000,0.000000)
\psline[linewidth=\AxesLineWidth,linecolor=GridColor](0.000000,0.100000)(0.012000,0.100000)
\psline[linewidth=\AxesLineWidth,linecolor=GridColor](0.000000,0.200000)(0.012000,0.200000)
\psline[linewidth=\AxesLineWidth,linecolor=GridColor](0.000000,0.300000)(0.012000,0.300000)
\psline[linewidth=\AxesLineWidth,linecolor=GridColor](0.000000,0.400000)(0.012000,0.400000)
\psline[linewidth=\AxesLineWidth,linecolor=GridColor](0.000000,0.500000)(0.012000,0.500000)
\psline[linewidth=\AxesLineWidth,linecolor=GridColor](0.000000,0.600000)(0.012000,0.600000)
\psline[linewidth=\AxesLineWidth,linecolor=GridColor](0.000000,0.700000)(0.012000,0.700000)
\psline[linewidth=\AxesLineWidth,linecolor=GridColor](0.000000,0.800000)(0.012000,0.800000)
\psline[linewidth=\AxesLineWidth,linecolor=GridColor](0.000000,0.900000)(0.012000,0.900000)
\psline[linewidth=\AxesLineWidth,linecolor=GridColor](0.000000,1.000000)(0.012000,1.000000)

{ \footnotesize 
\rput[t](0.000000,-0.015000){$0$}
\rput[t](0.100000,-0.015000){}
\rput[t](0.200000,-0.015000){}
\rput[t](0.300000,-0.015000){}
\rput[t](0.400000,-0.015000){}
\rput[t](0.500000,-0.015000){$0.5$}
\rput[t](0.600000,-0.015000){}
\rput[t](0.700000,-0.015000){}
\rput[t](0.800000,-0.015000){}
\rput[t](0.900000,-0.015000){}
\rput[t](1.000000,-0.015000){$1$}
\rput[r](-0.012000,0.000000){$0$}
\rput[r](-0.012000,0.100000){}
\rput[r](-0.012000,0.200000){}
\rput[r](-0.012000,0.300000){}
\rput[r](-0.012000,0.400000){}
\rput[r](-0.012000,0.500000){$0.5$}
\rput[r](-0.012000,0.600000){}
\rput[r](-0.012000,0.700000){}
\rput[r](-0.012000,0.800000){}
\rput[r](-0.012000,0.900000){}
\rput[r](-0.012000,1.000000){$1$}
} 

\pspolygon[linewidth=\AxesLineWidth](0.000000,0.000000)(1.000000,0.000000)(1.000000,1.000000)(0.000000,1.000000)(0.000000,0.000000)

{ \small 
\rput[b](0.500000,-0.193929){
\begin{tabular}{c}
Spatial frequency / Max.\ spatial frequency\\
\end{tabular}
}

\rput[t]{90}(-0.127143,0.500000){
\begin{tabular}{c}
Amplitude\\
\end{tabular}
}
} 

\newrgbcolor{color404.0182}{0  0  1}
\savedata{\mydata}[{
{0.000000,1.000000},{0.267000,0.664129},{0.427000,0.473332},{0.555000,0.331524},{0.662000,0.223543},
{0.751000,0.143452},{0.827000,0.084101},{0.891000,0.042486},{0.943000,0.016196},{0.983000,0.002654},
{0.998000,0.000107}
}]
\dataplot[plotstyle=line,linestyle=solid,linewidth=\LineWidth,linecolor=color404.0182]{\mydata}

\newrgbcolor{color405.0172}{0         0.5           0}
\savedata{\mydata}[{
{0.000000,1.000000},{0.091000,0.826281},{0.179000,0.674041},{0.263000,0.543169},{0.343000,0.431649},
{0.422000,0.334084},{0.498000,0.252004},{0.571000,0.184041},{0.642000,0.128164},{0.711000,0.083521},
{0.778000,0.049284},{0.843000,0.024649},{0.907000,0.008649},{0.970000,0.000900},{0.998000,0.000004},
}]
\dataplot[plotstyle=line,linestyle=solid,linewidth=\LineWidth,linecolor=color405.0172]{\mydata}

{ \small 
\rput[tr](0.976000,0.970000){%
\psshadowbox[framesep=0pt,linewidth=\AxesLineWidth]{\psframebox*{\begin{tabular}{l}
\Rnode{a1}{\hspace*{0.0ex}} \hspace*{0.3cm} \Rnode{a2}{~~OTF} \\
\Rnode{a3}{\hspace*{0.0ex}} \hspace*{0.3cm} \Rnode{a4}{~~Frequency Weighting} \\
\Rnode{a31}{\hspace*{0.0ex}} \hspace*{0.3cm} \Rnode{a42}{~~Function} \\
\end{tabular}}
\ncline[linestyle=solid,linewidth=\LineWidth,linecolor=color404.0182]{a1}{a2}
\ncline[linestyle=solid,linewidth=\LineWidth,linecolor=color405.0172]{a3}{a4}
}%
}%
} 

\end{pspicture}%
	\caption{One-dimensional normalized Optical Transfer Function (OTF) and frequency weighting function between direct component and maximum possible spatial frequency.}
	\label{fig:OTF_frequency_weighting}
\end{figure}

After the weighted projection of the residual onto all basis functions has been carried out, the one to be added to the model in the current iteration has to be determined. In the original Frequency Selective Extrapolation, the basis function $\varphi_{\left(k,l \right)}\left[m,n\right]$ is selected that minimizes the weighted distance between the residual and the according projection, that is to say the one which minimizes (\ref{eq:weighted_residual_iteration}). However, this criterion is not feasible for the considered reconstruction task. The problems that come up with this criterion are that ambiguities may arise due to the small number of available data points and therewith high-frequency basis functions that do not fit the actual content may get selected instead of a better fitting low frequency basis function. This leads to annoying ringing artifacts as shown in \cite{Schoberl2011a}. In order to cope with this, FSR uses a different selection criterion that is inspired by optical systems that have been used for acquiring the image. Looking at the Optical Transfer Function (OTF) as Fourier-transform of the point-spread function of a diffraction limited optical system, it can be seen that it is monotonically decreasing from the direct component to the maximally possible frequency. Figure \ref{fig:OTF_frequency_weighting} shows exemplarily the normalized OTF of the airy disk from a diffraction limited lens in one dimension. This transfer function also implies that high-frequency content gets attenuated more by the optical system than low-frequency one. This knowledge is now used to solve the ambiguities that may arise during the reconstruction process by introducing a multiplicative frequency weighting into the basis function selection process for taking the different probabilities of the individual basis functions into account.

\begin{figure}
\psfrag{k}[l][l]{$k$}
\psfrag{l}[r][l]{$l$}
\psfrag{sqrtvkl}[l][l][1][90]{$\sqrt{w_f\left[k,l\right]}$}
\psfrag{x01}[t][t]{$5$}%
\psfrag{x02}[t][t]{$10$}%
\psfrag{x03}[t][t]{$15$}%
\psfrag{x04}[t][t]{$20$}%
\psfrag{x05}[t][t]{$25$}%
\psfrag{x06}[t][t]{$30$}%
\psfrag{v01}[r][r]{$10$}%
\psfrag{v02}[r][r]{$20$}%
\psfrag{v03}[r][r]{$30$}%
\psfrag{z01}[r][r]{$0$}%
\psfrag{z02}[r][r]{$0.2$}%
\psfrag{z03}[r][r]{$0.4$}%
\psfrag{z04}[r][r]{$0.6$}%
\psfrag{z05}[r][r]{$0.8$}%
\psfrag{z06}[r][r]{$1$}%

\includegraphics[width=0.44\textwidth]{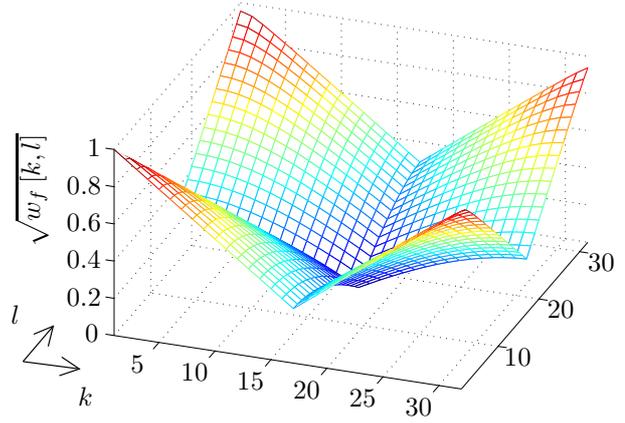}
\caption{Square root of frequency weighting function $w_f\left[k,l\right]$ for frequency-domain basis function selection.}
\label{fig:2d_root_weighting_function}
\end{figure}

Thus, the following frequency weighting function
\begin{equation}
w_f\left[k,l\right] = \left( 1 - \sqrt{2} \sqrt{\frac{\tilde{k}^2}{M^2} + \frac{\tilde{l}^2}{N^2}} \right)^2
\end{equation} 
is used during the selection process with the two substitutions $\tilde{k} = \frac{M}{2} - \left|k - \frac{M}{2}\right|$ and $\tilde{l} = \frac{N}{2} - \left|l-\frac{N}{2}\right|$ in order to allow for a compact representation. A one-dimensional plot of this frequency weighting function also is shown in \mbox{Figure \ref{fig:OTF_frequency_weighting}}. Defining $w_f\left[k,l\right]$ in this way has two advantages. First of all it can be regarded as an approximation of the OTF of an ideal system. An exact reproduction of the OTF is not necessary anyway since no optical system is ideal and typically, the exact parameters of the optical system are commonly not known. Of course, exploiting the exact OTF of the imaging system for the reconstruction would be beneficial. Nevertheless, the approximated OTF assures that high frequency ringing artifacts can be avoided by favoring low frequency basis functions over high frequency ones. However, if the latter ones are sufficiently dominant in the signal, they can still get selected as well and therewith allowing for the reconstruction of fine structures. In addition to this, as will be shown later, if the selection process is also carried out in the frequency domain, it results that the square root of $w_f\left[k,l\right]$ is used which is nothing else than a linearly decreasing function over the whole frequency support as shown in Figure \ref{fig:2d_root_weighting_function}. This assignment is quite convenient since it requires no additional parametrization and also represents well the different occurrence probabilities of the individual frequencies.

Using the frequency weighting function, the selection process results in
\[
\left(u,v\right) = \argmax_{\left(k,l \right)} \Bigg( \left|p_{\left(k,l \right)}^{\left(\nu\right) }\right|^2 w_f\left[k,l\right] \cdot\hspace{2cm}
\]\vspace{-0.5cm}
\begin{equation}
  \hspace{1.5cm}\sum_{\left(m,n\right)\in\mathcal{L}}  \varphi_{\left(k,l \right)}^\ast\left[m,n\right]w\left[m,n\right]\varphi_{\left(k,l \right)}\left[m,n\right]\Bigg)
\end{equation}
with $\left(u,v\right)$ being the index of the basis function to be added. This can also be expressed in the frequency domain by
\begin{eqnarray}
\left(u,v\right) &=& \argmax_{\left(k,l\right)} w_f\left[k,l\right] \frac{\left|R_w^{\left(\nu-1\right)}\left[k, l\right]\right|^2}{W\left[0,0\right]} \\
 &=& \argmax_{\left(k,l\right)} \sqrt{w_f\left[k,l\right]} \left|R_w^{\left(\nu-1\right)}\left[k, l\right]\right|
\end{eqnarray}
where the linearly decreasing $\sqrt{w_f\left[k,l\right]}$ can be used by exploiting the property that the $\argmax$ operation can be evaluated over the square root of the argument, here. Apparently, by further removing the constant $W\left[0,0\right]$ from the $\argmax$ calculation, the calculation of the basis function to be added in iteration $\nu$ can be expressed very compactly and calculated quite efficiently. 

After the basis function to be added has been selected, its corresponding weight has to be estimated. This is achieved by
\begin{equation}
 \hat{c}_{\left(u,v \right)}^{\left(\nu\right)} = \gamma p_{\left(u,v \right)}^{\left(\nu\right)}  = \gamma\frac{R_w^{\left(\nu-1\right)}\left[u,v\right]}{W\left[0,0\right]}
\end{equation}
where $\gamma$ is the orthogonality deficiency compensation factor used for obtaining a stable estimation and reducing interference between the different basis functions. This compensation is necessary since the basis function are not orthogonal anymore, if evaluated together with the weighting function $w\left[m,n\right]$. As shown in \cite{Seiler2008}, a constant $\gamma$ could be used as a good approximation for the elaborate compensation of the orthogonality deficiency which is proposed in \cite{Seiler2007}.

After basis function selection and weight estimation, all samples of the model generated so far are updated according to
\begin{equation}
  g^{\left(\nu\right)}\left[m,n\right] = g^{\left(\nu-1\right)}\left[m,n\right] + \hat{c}_{\left(u,v \right)}^{\left(\nu\right)} \varphi_{\left(u,v \right)}\left[m,n\right]
\end{equation}
which can also be expressed quite efficiently in the frequency domain by
\begin{equation} 
G^{\left(\nu\right)}\left[u,v\right] = G^{\left(\nu-1\right)}\left[u,v\right] + MN\hat{c}_{\left(u,v \right)}^{\left(\nu\right)}
\end{equation}
where only the amplitude of the selected basis function with index $\left(u,v\right)$ has to be changed. Besides the model, also the residual has to be updated by the selected basis function
\begin{equation}
r^{\left(\nu\right)}\left[m,n\right] = r^{\left(\nu-1\right)}\left[m,n\right] - \hat{c}_{\left(u,v \right)}^{\left(\nu\right)} \varphi_{\left(u,v \right)} \left[m,n\right].
\end{equation}
This step can be expressed in the frequency domain as well by
\begin{equation}
 R_w^{\left(\nu\right)} \left[k,l\right] = R_w^{\left(\nu-1\right)} \left[k,l\right] - \hat{c}_{\left(u,v \right)}^{\left(\nu\right)} W\left[k-u, l-v\right], \forall\left(k,l\right).
\end{equation}
After the update, the model generation process proceeds to the next iteration where another basis function is selected and its weight is estimated. These steps are repeated until a predefined number of iterations is reached. As all operations within the iteration can be expressed in the frequency domain, it is advisable to carry out the whole model generation in the frequency domain and therewith avoiding the explicit calculation of the projection coefficients. In order to achieve this, only a transform into the frequency domain at the beginning and one back into the spatial domain after the modeling has finished are necessary.

Finally, the real-valued part of the samples from the model that correspond to unknown samples of the currently considered block are taken as estimate for the content of the block and therewith replacing the unknown samples. After this, the reconstruction process proceeds to the next block while all the samples of the just finished block can be used for improving the reconstruction of the neighboring, yet unprocessed blocks.


\section{Relationship to Compressed Sensing}\label{sec:compressed_sensing} 
Even though the above outlined FSR has a completely different background and has developed from a different origin, it can also be interpreted in the Compressed Sensing (CS) framework \cite{Candes2006, Donoho2006}. CS is a concept which came up some years ago and which states that the number of samples that have to be acquired from a signal may be much smaller compared to the classical Shannon-Nyquist theorem \cite{Unser2000}, as long as a sparse representation of the signal exists. In order to achieve this, the sparsity of the signal is exploited directly during the acquisition process by measuring multiple linear combinations of the sparse components. 

In order to describe the subsampling process and the resampling by FSR within the CS framework, the desired signal $f\left[m,n\right]$ on the regular grid is regarded in vectorized form as
\begin{equation}
\label{eq:CS_signal_definition}
\ve{f}= \ve{\Phi} \ve{c}
\end{equation}
with $\ve{\Phi}$ being the matrix of the vectorized basis functions and $\ve{c}$ being the sparse vector of expansion coefficients. As vector $\ve{f}$ contains as many samples as $f\left[m,n\right]$ it is of length
\begin{equation}
L_{\ve{f}} = MN.
\end{equation}
The matrix $\ve{\Phi}$ contains all possible basis functions and therewith forms a complete Fourier basis. Accordingly, matrix $\ve{\Phi}$ is of size $\left(MN\right)\times\left(MN\right)$. Using (\ref{eq:CS_signal_definition}), the generation process for the subsampled signal from (\ref{eq:non_reg_generation_block}) can be written according to the typical CS notation by
\begin{eqnarray}
\ve{f}_{\nr} &=& \ve{Q}\ve{f} \\
\label{eq:CS_measurement} &=& \ve{Q}\ve{\Phi} \ve{c}
\end{eqnarray}
with $\ve{f}_{\nr}$ being the available subsampled signal $f_{\nr} \left[m,n\right]$ written in vectorized form. In this context, the entries, where $q \left[m,n\right]$ is equal to zero are not part of $\ve{f}_{\nr}$. Thus, the vector $\ve{f}_{\nr}$ is of length 
\begin{equation}
L_{\ve{f}_{\nr}}  = \sum_{\forall\left(m,n\right)} q\left[m,n\right].
\end{equation}
The matrix $\ve{Q}$ is the subsampling matrix which is of size $L_{\ve{f}_{\nr}}\times L_{\ve{f}}$ and results from $q\left[m,n\right]$ and determines the available samples after the non-regular subsampling. Since the subsampling process performs a sample-wise mapping from the pixels from $\ve{f}$ to $\ve{f}_{\nr}$, every row contains only zeros except for a single one. All the samples which get lost by the subsampling process are represented by columns completely full of zeros. Hence $\ve{Q}$ has a form like
\begin{equation}
\ve{Q} = \left(
\begin{array}{ccccccccc}
0 & 0 & 1 & 0 & 0 & 0 & 0 & \ldots & 0 \\
0 & 0& 0 & 0 & 0 & 0 & 0 & \ldots & 1 \\
0 & 0& 0 & 0 & 0 & 1 & 0 & \ldots & 0 \\
&&& \vdots &&&&& \vdots \\
1 & 0& 0 & 0 & 0 & 0 & 0 & \ldots & 0 \\
\end{array}
\right) .
\end{equation}
It has to be noted that matrix $\ve{Q}$ which acts as measurement matrix is a very sparse matrix and therewith differs from the matrices most often used in CS. However, in combination with sparsifying matrix $\ve{\Phi}$, the product of $\ve{Q}\ve{\Phi}$ in (\ref{eq:CS_measurement}) can be regarded as the matrix for obtaining the linear measurements of the sparse coefficient vector $\ve{c}$.

Similar to CS, the objective of the proposed FSR is to estimate the sparse coefficient vector $\ve{c}$ based on the vector $\ve{f}_{\nr}$ and the measurement matrix $\ve{Q}$ together with the transform matrix $\ve{\Phi}$. For this, FSR generates the model $g\left[m,n\right]$ which can be written as
\begin{equation}
\ve{g}= \ve{\Phi} \hat{\ve{c}}
\end{equation}
with the coefficient vector $\hat{\ve{c}}$ to be determined by the sparse modeling of FSR.

For generating the coefficient vector $\hat{\ve{c}}$, the operations FSR carries out are related to the ones from Matching Pursuits (MP) \cite{Mallat1993}. Therewith, FSR could be regarded as to belong to the class of so-called greedy recovery algorithms \cite{Eldar2012}. In doing so, the generation of a sparse solution is assured since only a limited number of iterations is carried out. Thus, the number of non-zero elements in  $\hat{\ve{c}}$ is maximally as large as the number of iterations. As a basis function can get selected multiple times, it might also be smaller.

However, even though the modeling is related to MP, the proposed FSR significantly differs from it in several points. First of all, MP minimizes the approximation energy with respect to the selected basis function in every iteration \cite{Mallat1993, Eldar2012}. In contrast to this, FSR uses the weighted residual energy 
\begin{equation}
E_w = \left(\bar{\ve{f}}_{\nr} - \ve{Q}\ve{\Phi}\hat{\ve{c}}\right)^\mathrm{H} \ve{W} \left(\bar{\ve{f}}_{\nr} - \ve{Q}\ve{\Phi}\hat{\ve{c}}\right)
\end{equation}
only as an intermediate step for selecting the basis functions during the iterations and estimating its weight. Here, $\ve{W}$ is a diagonal matrix with the non-zero entries of $w\left[m,n\right]$ on the main diagonal. Unlike MP, by applying the orthogonality deficiency compensation, FSR does not maximally reduce the energy term in every iteration with respect to the selected basis function, but rather only a fraction of the basis function is added to the model in every iteration. 

On top of this, the spatial weighting function $w\left[m,n\right]$ is not part of the original MP and also the frequency weighting is not used, there. Instead, FSR exploits the prior knowledge that in natural images high frequencies are less likely to occur than low ones by applying the frequency weighting $w_f\left[k,l\right]$ during the selection process. In CS theory, there also exist related approaches like \cite{Masood2013, Chi2013} which make use of some prior knowledge for recovering the signal. However, the prior knowledge is exploited in different ways, there. For exploiting a prior knowledge about the distribution of the non-zero entries in the coefficient vector, \cite{Masood2013} iteratively estimates the support of the coefficient vector. In doing so, the dimensionality of the reconstruction problem can be reduced leading to an improved reconstruction quality. In \cite{Chi2013}, a so called oblique projection is used for the reconstruction. For this, not the measurement matrix is used for the reconstruction by MP, but rather a second matrix is generated and used, in order to match the probability distribution of the signal support more. 

In contrast to this, the proposed FSR exploits the prior knowledge by carrying out the weighted projection and applying the frequency weighting during the selection process. It has to be further noted, that, since FSR only operates on small areas of the image, it is related most to block-wise CS algorithms \cite{Mun2009}. Even though there exist CS algorithms like \cite{Gutierrez2014} which use overlapping blocks, as well, the proposed FSR goes beyond this by reusing already reconstructed samples from one block to support the recovery of the next block and especially making the processing order adaptive to the local density of the available samples. 

Altogether, FSR can indeed be interpreted within the CS framework. Despite having a different origin, it combines many techniques which have also been developed independently within the CS framework, like the iterative modeling, the exploitation of prior knowledge, and the block processing. Nevertheless, FSR also uses different concepts like the spatial weighting, the orthogonality deficiency compensation, or the optimized processing order in combination with the reuse of already reconstructed blocks which are not common in CS. The combination of the effective modeling of the signal based on the available samples within each block together with the optimized processing order which can account for locally varying densities allows for a very high reconstruction quality, as will be shown in the next section. Furthermore, an additional discussion is provided, analyzing the performance of FSR, especially in comparison with other state-of-the-art reconstruction algorithms, at the end of the next section.


\section{Simulations and Results}\label{sec:simulations} 
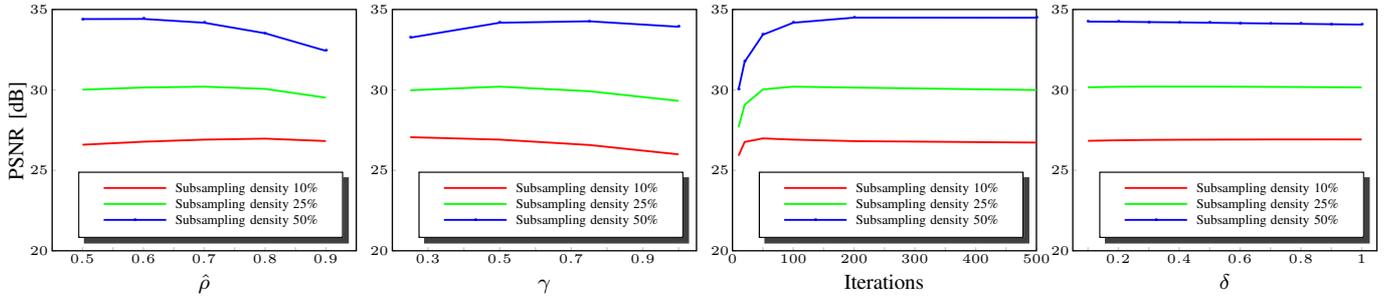
\begin{figure*} 
	\centering
	\begin{minipage}{0.24\textwidth}



\providelength{\AxesLineWidth}       \setlength{\AxesLineWidth}{0.5pt}
\providelength{\plotwidth}           \setlength{\plotwidth}{4cm} 
\providelength{\LineWidth}           \setlength{\LineWidth}{0.7pt}
\providelength{\MarkerSize}          \setlength{\MarkerSize}{4pt}
\newrgbcolor{GridColor}{0.8 0.8 0.8}

\psset{xunit=2.000000\plotwidth,yunit=0.053333\plotwidth}
\begin{pspicture}(0.375000,17.187500)(0.950000,35.000000)


\psline[linewidth=\AxesLineWidth,linecolor=GridColor](0.450000,20.000000)(0.450000,20.225000)
\psline[linewidth=\AxesLineWidth,linecolor=GridColor](0.500000,20.000000)(0.500000,20.225000)
\psline[linewidth=\AxesLineWidth,linecolor=GridColor](0.550000,20.000000)(0.550000,20.225000)
\psline[linewidth=\AxesLineWidth,linecolor=GridColor](0.600000,20.000000)(0.600000,20.225000)
\psline[linewidth=\AxesLineWidth,linecolor=GridColor](0.650000,20.000000)(0.650000,20.225000)
\psline[linewidth=\AxesLineWidth,linecolor=GridColor](0.700000,20.000000)(0.700000,20.225000)
\psline[linewidth=\AxesLineWidth,linecolor=GridColor](0.750000,20.000000)(0.750000,20.225000)
\psline[linewidth=\AxesLineWidth,linecolor=GridColor](0.800000,20.000000)(0.800000,20.225000)
\psline[linewidth=\AxesLineWidth,linecolor=GridColor](0.850000,20.000000)(0.850000,20.225000)
\psline[linewidth=\AxesLineWidth,linecolor=GridColor](0.900000,20.000000)(0.900000,20.225000)
\psline[linewidth=\AxesLineWidth,linecolor=GridColor](0.950000,20.000000)(0.950000,20.225000)
\psline[linewidth=\AxesLineWidth,linecolor=GridColor](0.450000,20.000000)(0.456000,20.000000)
\psline[linewidth=\AxesLineWidth,linecolor=GridColor](0.450000,25.000000)(0.456000,25.000000)
\psline[linewidth=\AxesLineWidth,linecolor=GridColor](0.450000,30.000000)(0.456000,30.000000)
\psline[linewidth=\AxesLineWidth,linecolor=GridColor](0.450000,35.000000)(0.456000,35.000000)

{ \tiny 
\rput[t](0.450000,19.775000){}
\rput[t](0.500000,19.775000){$0.5$}
\rput[t](0.550000,19.775000){}
\rput[t](0.600000,19.775000){$0.6$}
\rput[t](0.650000,19.775000){}
\rput[t](0.700000,19.775000){$0.7$}
\rput[t](0.750000,19.775000){}
\rput[t](0.800000,19.775000){$0.8$}
\rput[t](0.850000,19.775000){}
\rput[t](0.900000,19.775000){$0.9$}
\rput[t](0.950000,19.775000){}
\rput[r](0.444000,20.000000){$20$}
\rput[r](0.444000,25.000000){$25$}
\rput[r](0.444000,30.000000){$30$}
\rput[r](0.444000,35.000000){$35$}
} 

\pspolygon[linewidth=\AxesLineWidth](0.450000,20.000000)(0.950000,20.000000)(0.950000,35.000000)(0.450000,35.000000)(0.450000,20.000000)

{ \footnotesize 
\rput[b](0.700000,17.187500){
\begin{tabular}{c}
$\hat{\rho}$\\
\end{tabular}
}

\rput[t]{90}(0.375000,27.500000){
\begin{tabular}{c}
PSNR [dB]\\
\end{tabular}
}
} 

\newrgbcolor{color268.0052}{1  0  0}
\savedata{\mydata}[{
{0.500000,26.592299},{0.600000,26.779324},{0.700000,26.909037},{0.800000,26.970490},{0.900000,26.821224},
}]
\dataplot[plotstyle=line,showpoints=true,dotstyle=+,dotsize=\MarkerSize,linestyle=solid,linewidth=\LineWidth,linecolor=color268.0052]{\mydata}

\newrgbcolor{color269.0052}{0  1  0}
\savedata{\mydata}[{
{0.500000,30.015655},{0.600000,30.148604},{0.700000,30.201875},{0.800000,30.063965},{0.900000,29.519126},
}]
\dataplot[plotstyle=line,showpoints=true,dotstyle=asterisk,dotsize=\MarkerSize,linestyle=solid,linewidth=\LineWidth,linecolor=color269.0052]{\mydata}

\newrgbcolor{color270.0052}{0  0  1}
\savedata{\mydata}[{
{0.500000,34.403166},{0.600000,34.412073},{0.700000,34.175686},{0.800000,33.523228},{0.900000,32.429532},
}]
\dataplot[plotstyle=line,showpoints=true,dotstyle=x,dotsize=\MarkerSize,linestyle=solid,linewidth=\LineWidth,linecolor=color270.0052]{\mydata}

{ \tiny 
\rput[br](0.938000,20.450000){%
\psshadowbox[framesep=0pt,linewidth=\AxesLineWidth]{\psframebox*{\begin{tabular}{l}
\Rnode{a1}{\hspace*{0.0ex}} \hspace*{0.7cm} \Rnode{a2}{~~Subsampling density 10\%} \\
\Rnode{a3}{\hspace*{0.0ex}} \hspace*{0.7cm} \Rnode{a4}{~~Subsampling density 25\%} \\
\Rnode{a5}{\hspace*{0.0ex}} \hspace*{0.7cm} \Rnode{a6}{~~Subsampling density 50\%} \\
\end{tabular}}
\ncline[linestyle=solid,linewidth=\LineWidth,linecolor=color268.0052]{a1}{a2} \ncput{\psdot[dotstyle=+,dotsize=\MarkerSize,linecolor=color268.0052]}
\ncline[linestyle=solid,linewidth=\LineWidth,linecolor=color269.0052]{a3}{a4} \ncput{\psdot[dotstyle=asterisk,dotsize=\MarkerSize,linecolor=color269.0052]}
\ncline[linestyle=solid,linewidth=\LineWidth,linecolor=color270.0052]{a5}{a6} \ncput{\psdot[dotstyle=x,dotsize=\MarkerSize,linecolor=color270.0052]}
}%
}%
} 

\end{pspicture}%
	\end{minipage}
	\begin{minipage}{0.24\textwidth}



\providelength{\AxesLineWidth}       \setlength{\AxesLineWidth}{0.5pt}
\providelength{\plotwidth}           \setlength{\plotwidth}{4cm} 
\providelength{\LineWidth}           \setlength{\LineWidth}{0.7pt}
\providelength{\MarkerSize}          \setlength{\MarkerSize}{4pt}
\newrgbcolor{GridColor}{0.8 0.8 0.8}

\psset{xunit=1.176471\plotwidth,yunit=0.053333\plotwidth}
\begin{pspicture}(0.072500,17.187500)(1.050000,35.000000)


\psline[linewidth=\AxesLineWidth,linecolor=GridColor](0.200000,20.000000)(0.200000,20.225000)
\psline[linewidth=\AxesLineWidth,linecolor=GridColor](0.300000,20.000000)(0.300000,20.225000)
\psline[linewidth=\AxesLineWidth,linecolor=GridColor](0.400000,20.000000)(0.400000,20.225000)
\psline[linewidth=\AxesLineWidth,linecolor=GridColor](0.500000,20.000000)(0.500000,20.225000)
\psline[linewidth=\AxesLineWidth,linecolor=GridColor](0.600000,20.000000)(0.600000,20.225000)
\psline[linewidth=\AxesLineWidth,linecolor=GridColor](0.700000,20.000000)(0.700000,20.225000)
\psline[linewidth=\AxesLineWidth,linecolor=GridColor](0.800000,20.000000)(0.800000,20.225000)
\psline[linewidth=\AxesLineWidth,linecolor=GridColor](0.900000,20.000000)(0.900000,20.225000)
\psline[linewidth=\AxesLineWidth,linecolor=GridColor](1.000000,20.000000)(1.000000,20.225000)
\psline[linewidth=\AxesLineWidth,linecolor=GridColor](0.200000,20.000000)(0.210200,20.000000)
\psline[linewidth=\AxesLineWidth,linecolor=GridColor](0.200000,25.000000)(0.210200,25.000000)
\psline[linewidth=\AxesLineWidth,linecolor=GridColor](0.200000,30.000000)(0.210200,30.000000)
\psline[linewidth=\AxesLineWidth,linecolor=GridColor](0.200000,35.000000)(0.210200,35.000000)

{ \tiny 
\rput[t](0.200000,19.775000){}
\rput[t](0.300000,19.775000){$0.3$}
\rput[t](0.400000,19.775000){}
\rput[t](0.500000,19.775000){$0.5$}
\rput[t](0.600000,19.775000){}
\rput[t](0.700000,19.775000){$0.7$}
\rput[t](0.800000,19.775000){}
\rput[t](0.900000,19.775000){$0.9$}
\rput[t](1.000000,19.775000){}
\rput[r](0.189800,20.000000){$20$}
\rput[r](0.189800,25.000000){$25$}
\rput[r](0.189800,30.000000){$30$}
\rput[r](0.189800,35.000000){$35$}
} 

\pspolygon[linewidth=\AxesLineWidth](0.200000,20.000000)(1.050000,20.000000)(1.050000,35.000000)(0.200000,35.000000)(0.200000,20.000000)

{ \footnotesize 
\rput[b](0.625000,17.187500){
\begin{tabular}{c}
$\gamma$\\
\end{tabular}
}

\rput[t]{90}(0.072500,27.500000){
\begin{tabular}{c}
\end{tabular}
}
} 

\newrgbcolor{color268.0059}{1  0  0}
\savedata{\mydata}[{
{0.250000,27.061130},{0.500000,26.909037},{0.750000,26.576887},{1.000000,25.993496}
}]
\dataplot[plotstyle=line,showpoints=true,dotstyle=+,dotsize=\MarkerSize,linestyle=solid,linewidth=\LineWidth,linecolor=color268.0059]{\mydata}

\newrgbcolor{color269.0054}{0  1  0}
\savedata{\mydata}[{
{0.250000,29.979186},{0.500000,30.201875},{0.750000,29.918504},{1.000000,29.318775}
}]
\dataplot[plotstyle=line,showpoints=true,dotstyle=asterisk,dotsize=\MarkerSize,linestyle=solid,linewidth=\LineWidth,linecolor=color269.0054]{\mydata}

\newrgbcolor{color270.0054}{0  0  1}
\savedata{\mydata}[{
{0.250000,33.249184},{0.500000,34.175686},{0.750000,34.266079},{1.000000,33.926342}
}]
\dataplot[plotstyle=line,showpoints=true,dotstyle=x,dotsize=\MarkerSize,linestyle=solid,linewidth=\LineWidth,linecolor=color270.0054]{\mydata}

{ \tiny 
\rput[br](1.029600,20.450000){%
\psshadowbox[framesep=0pt,linewidth=\AxesLineWidth]{\psframebox*{\begin{tabular}{l}
\Rnode{a1}{\hspace*{0.0ex}} \hspace*{0.7cm} \Rnode{a2}{~~Subsampling density 10\%} \\
\Rnode{a3}{\hspace*{0.0ex}} \hspace*{0.7cm} \Rnode{a4}{~~Subsampling density 25\%} \\
\Rnode{a5}{\hspace*{0.0ex}} \hspace*{0.7cm} \Rnode{a6}{~~Subsampling density 50\%} \\
\end{tabular}}
\ncline[linestyle=solid,linewidth=\LineWidth,linecolor=color268.0059]{a1}{a2} \ncput{\psdot[dotstyle=+,dotsize=\MarkerSize,linecolor=color268.0059]}
\ncline[linestyle=solid,linewidth=\LineWidth,linecolor=color269.0054]{a3}{a4} \ncput{\psdot[dotstyle=asterisk,dotsize=\MarkerSize,linecolor=color269.0054]}
\ncline[linestyle=solid,linewidth=\LineWidth,linecolor=color270.0054]{a5}{a6} \ncput{\psdot[dotstyle=x,dotsize=\MarkerSize,linecolor=color270.0054]}
}%
}%
} 

\end{pspicture}%
	\end{minipage}
	\begin{minipage}{0.24\textwidth}



\providelength{\AxesLineWidth}       \setlength{\AxesLineWidth}{0.5pt}
\providelength{\plotwidth}           \setlength{\plotwidth}{4cm} 
\providelength{\LineWidth}           \setlength{\LineWidth}{0.7pt}
\providelength{\MarkerSize}          \setlength{\MarkerSize}{4pt}
\newrgbcolor{GridColor}{0.8 0.8 0.8}

\psset{xunit=0.002000\plotwidth,yunit=0.053333\plotwidth}
\begin{pspicture}(-75.000000,17.187500)(500.000000,35.000000)


\psline[linewidth=\AxesLineWidth,linecolor=GridColor](0.000000,20.000000)(0.000000,20.225000)
\psline[linewidth=\AxesLineWidth,linecolor=GridColor](50.000000,20.000000)(50.000000,20.225000)
\psline[linewidth=\AxesLineWidth,linecolor=GridColor](100.000000,20.000000)(100.000000,20.225000)
\psline[linewidth=\AxesLineWidth,linecolor=GridColor](150.000000,20.000000)(150.000000,20.225000)
\psline[linewidth=\AxesLineWidth,linecolor=GridColor](200.000000,20.000000)(200.000000,20.225000)
\psline[linewidth=\AxesLineWidth,linecolor=GridColor](250.000000,20.000000)(250.000000,20.225000)
\psline[linewidth=\AxesLineWidth,linecolor=GridColor](300.000000,20.000000)(300.000000,20.225000)
\psline[linewidth=\AxesLineWidth,linecolor=GridColor](350.000000,20.000000)(350.000000,20.225000)
\psline[linewidth=\AxesLineWidth,linecolor=GridColor](400.000000,20.000000)(400.000000,20.225000)
\psline[linewidth=\AxesLineWidth,linecolor=GridColor](450.000000,20.000000)(450.000000,20.225000)
\psline[linewidth=\AxesLineWidth,linecolor=GridColor](500.000000,20.000000)(500.000000,20.225000)
\psline[linewidth=\AxesLineWidth,linecolor=GridColor](0.000000,20.000000)(6.000000,20.000000)
\psline[linewidth=\AxesLineWidth,linecolor=GridColor](0.000000,25.000000)(6.000000,25.000000)
\psline[linewidth=\AxesLineWidth,linecolor=GridColor](0.000000,30.000000)(6.000000,30.000000)
\psline[linewidth=\AxesLineWidth,linecolor=GridColor](0.000000,35.000000)(6.000000,35.000000)

{ \tiny 
\rput[t](0.000000,19.775000){$0$}
\rput[t](50.000000,19.775000){}
\rput[t](100.000000,19.775000){$100$}
\rput[t](150.000000,19.775000){}
\rput[t](200.000000,19.775000){$200$}
\rput[t](250.000000,19.775000){}
\rput[t](300.000000,19.775000){$300$}
\rput[t](350.000000,19.775000){}
\rput[t](400.000000,19.775000){$400$}
\rput[t](450.000000,19.775000){}
\rput[t](500.000000,19.775000){$500$}
\rput[r](-6.000000,20.000000){$20$}
\rput[r](-6.000000,25.000000){$25$}
\rput[r](-6.000000,30.000000){$30$}
\rput[r](-6.000000,35.000000){$35$}
} 

\pspolygon[linewidth=\AxesLineWidth](0.000000,20.000000)(500.000000,20.000000)(500.000000,35.000000)(0.000000,35.000000)(0.000000,20.000000)

{ \footnotesize 
\rput[b](250.000000,17.187500){
\begin{tabular}{c}
Iterations\\
\end{tabular}
}

\rput[t]{90}(-75.000000,27.500000){
\begin{tabular}{c}
\end{tabular}
}
} 

\newrgbcolor{color276.0063}{1  0  0}
\savedata{\mydata}[{
{10.000000,25.913122},{20.000000,26.765152},{50.000000,26.985328},{100.000000,26.909037},{200.000000,26.815971},
{500.000000,26.726338}
}]
\dataplot[plotstyle=line,showpoints=true,dotstyle=+,dotsize=\MarkerSize,linestyle=solid,linewidth=\LineWidth,linecolor=color276.0063]{\mydata}

\newrgbcolor{color277.0057}{0  1  0}
\savedata{\mydata}[{
{10.000000,27.681066},{20.000000,29.072173},{50.000000,30.030761},{100.000000,30.201875},{200.000000,30.143962},
{500.000000,29.993296}
}]
\dataplot[plotstyle=line,showpoints=true,dotstyle=asterisk,dotsize=\MarkerSize,linestyle=solid,linewidth=\LineWidth,linecolor=color277.0057]{\mydata}

\newrgbcolor{color278.0059}{0  0  1}
\savedata{\mydata}[{
{10.000000,30.033599},{20.000000,31.751291},{50.000000,33.431790},{100.000000,34.175686},{200.000000,34.492670},
{500.000000,34.487808}
}]
\dataplot[plotstyle=line,showpoints=true,dotstyle=x,dotsize=\MarkerSize,linestyle=solid,linewidth=\LineWidth,linecolor=color278.0059]{\mydata}

{ \tiny 
\rput[br](488.000000,20.450000){%
\psshadowbox[framesep=0pt,linewidth=\AxesLineWidth]{\psframebox*{\begin{tabular}{l}
\Rnode{a1}{\hspace*{0.0ex}} \hspace*{0.7cm} \Rnode{a2}{~~Subsampling density 10\%} \\
\Rnode{a3}{\hspace*{0.0ex}} \hspace*{0.7cm} \Rnode{a4}{~~Subsampling density 25\%} \\
\Rnode{a5}{\hspace*{0.0ex}} \hspace*{0.7cm} \Rnode{a6}{~~Subsampling density 50\%} \\
\end{tabular}}
\ncline[linestyle=solid,linewidth=\LineWidth,linecolor=color276.0063]{a1}{a2} \ncput{\psdot[dotstyle=+,dotsize=\MarkerSize,linecolor=color276.0063]}
\ncline[linestyle=solid,linewidth=\LineWidth,linecolor=color277.0057]{a3}{a4} \ncput{\psdot[dotstyle=asterisk,dotsize=\MarkerSize,linecolor=color277.0057]}
\ncline[linestyle=solid,linewidth=\LineWidth,linecolor=color278.0059]{a5}{a6} \ncput{\psdot[dotstyle=x,dotsize=\MarkerSize,linecolor=color278.0059]}
}%
}%
} 

\end{pspicture}%
	\end{minipage}
	\begin{minipage}{0.24\textwidth}



\providelength{\AxesLineWidth}       \setlength{\AxesLineWidth}{0.5pt}
\providelength{\plotwidth}           \setlength{\plotwidth}{4cm} 
\providelength{\LineWidth}           \setlength{\LineWidth}{0.7pt}
\providelength{\MarkerSize}          \setlength{\MarkerSize}{4pt}
\newrgbcolor{GridColor}{0.8 0.8 0.8}

\psset{xunit=1.000000\plotwidth,yunit=0.053333\plotwidth}
\begin{pspicture}(-0.100000,17.187500)(1.050000,35.000000)


\psline[linewidth=\AxesLineWidth,linecolor=GridColor](0.100000,20.000000)(0.100000,20.225000)
\psline[linewidth=\AxesLineWidth,linecolor=GridColor](0.200000,20.000000)(0.200000,20.225000)
\psline[linewidth=\AxesLineWidth,linecolor=GridColor](0.300000,20.000000)(0.300000,20.225000)
\psline[linewidth=\AxesLineWidth,linecolor=GridColor](0.400000,20.000000)(0.400000,20.225000)
\psline[linewidth=\AxesLineWidth,linecolor=GridColor](0.500000,20.000000)(0.500000,20.225000)
\psline[linewidth=\AxesLineWidth,linecolor=GridColor](0.600000,20.000000)(0.600000,20.225000)
\psline[linewidth=\AxesLineWidth,linecolor=GridColor](0.700000,20.000000)(0.700000,20.225000)
\psline[linewidth=\AxesLineWidth,linecolor=GridColor](0.800000,20.000000)(0.800000,20.225000)
\psline[linewidth=\AxesLineWidth,linecolor=GridColor](0.900000,20.000000)(0.900000,20.225000)
\psline[linewidth=\AxesLineWidth,linecolor=GridColor](1.000000,20.000000)(1.000000,20.225000)
\psline[linewidth=\AxesLineWidth,linecolor=GridColor](0.050000,20.000000)(0.062000,20.000000)
\psline[linewidth=\AxesLineWidth,linecolor=GridColor](0.050000,25.000000)(0.062000,25.000000)
\psline[linewidth=\AxesLineWidth,linecolor=GridColor](0.050000,30.000000)(0.062000,30.000000)
\psline[linewidth=\AxesLineWidth,linecolor=GridColor](0.050000,35.000000)(0.062000,35.000000)

{ \tiny 
\rput[t](0.100000,19.775000){}
\rput[t](0.200000,19.775000){$0.2$}
\rput[t](0.300000,19.775000){}
\rput[t](0.400000,19.775000){$0.4$}
\rput[t](0.500000,19.775000){}
\rput[t](0.600000,19.775000){$0.6$}
\rput[t](0.700000,19.775000){}
\rput[t](0.800000,19.775000){$0.8$}
\rput[t](0.900000,19.775000){}
\rput[t](1.000000,19.775000){$1$}
\rput[r](0.038000,20.000000){$20$}
\rput[r](0.038000,25.000000){$25$}
\rput[r](0.038000,30.000000){$30$}
\rput[r](0.038000,35.000000){$35$}
} 

\pspolygon[linewidth=\AxesLineWidth](0.050000,20.000000)(1.050000,20.000000)(1.050000,35.000000)(0.050000,35.000000)(0.050000,20.000000)

{ \footnotesize 
\rput[b](0.550000,17.187500){
\begin{tabular}{c}
$\delta$\\
\end{tabular}
}

\rput[t]{90}(-0.100000,27.500000){
\begin{tabular}{c}
\end{tabular}
}
} 

\newrgbcolor{color272.0061}{1  0  0}
\savedata{\mydata}[{
{0.100000,26.834322},{0.200000,26.869087},{0.300000,26.890368},{0.400000,26.901778},{0.500000,26.909037},
{0.600000,26.914326},{0.700000,26.921177},{0.800000,26.920826},{0.900000,26.922037},{1.000000,26.920692},
}]
\dataplot[plotstyle=line,showpoints=true,dotstyle=+,dotsize=\MarkerSize,linestyle=solid,linewidth=\LineWidth,linecolor=color272.0061]{\mydata}

\newrgbcolor{color273.0056}{0  1  0}
\savedata{\mydata}[{
{0.100000,30.161682},{0.200000,30.195403},{0.300000,30.205276},{0.400000,30.204611},{0.500000,30.201875},
{0.600000,30.196209},{0.700000,30.187854},{0.800000,30.179421},{0.900000,30.168271},{1.000000,30.156294},
}]
\dataplot[plotstyle=line,showpoints=true,dotstyle=asterisk,dotsize=\MarkerSize,linestyle=solid,linewidth=\LineWidth,linecolor=color273.0056]{\mydata}

\newrgbcolor{color274.0057}{0  0  1}
\savedata{\mydata}[{
{0.100000,34.245335},{0.200000,34.235585},{0.300000,34.217201},{0.400000,34.198319},{0.500000,34.175686},
{0.600000,34.151190},{0.700000,34.129320},{0.800000,34.106025},{0.900000,34.085106},{1.000000,34.060262},
}]
\dataplot[plotstyle=line,showpoints=true,dotstyle=x,dotsize=\MarkerSize,linestyle=solid,linewidth=\LineWidth,linecolor=color274.0057]{\mydata}

{ \tiny 
\rput[br](1.026000,20.450000){%
\psshadowbox[framesep=0pt,linewidth=\AxesLineWidth]{\psframebox*{\begin{tabular}{l}
\Rnode{a1}{\hspace*{0.0ex}} \hspace*{0.7cm} \Rnode{a2}{~~Subsampling density 10\%} \\
\Rnode{a3}{\hspace*{0.0ex}} \hspace*{0.7cm} \Rnode{a4}{~~Subsampling density 25\%} \\
\Rnode{a5}{\hspace*{0.0ex}} \hspace*{0.7cm} \Rnode{a6}{~~Subsampling density 50\%} \\
\end{tabular}}
\ncline[linestyle=solid,linewidth=\LineWidth,linecolor=color272.0061]{a1}{a2} \ncput{\psdot[dotstyle=+,dotsize=\MarkerSize,linecolor=color272.0061]}
\ncline[linestyle=solid,linewidth=\LineWidth,linecolor=color273.0056]{a3}{a4} \ncput{\psdot[dotstyle=asterisk,dotsize=\MarkerSize,linecolor=color273.0056]}
\ncline[linestyle=solid,linewidth=\LineWidth,linecolor=color274.0057]{a5}{a6} \ncput{\psdot[dotstyle=x,dotsize=\MarkerSize,linecolor=color274.0057]}
}%
}%
} 

\end{pspicture}%
	\end{minipage}
	\caption{Average reconstruction quality in PSNR for Kodak test data base for different subsampling densities and varying weighting function decay factors $\hat{\rho}$, orthogonality deficiency compensation factors $\gamma$, number of iterations, and weighting factors $\delta$ of already reconstructed areas.}
	\label{fig:kodak_psnr_over_parameters}
\end{figure*}

In order to illustrate how well FSR can be used for resampling images from a non-regular subset of pixel positions to a regular two-dimensional grid, this section is intended to show simulation results and a comparison to alternative reconstruction algorithms. First of all, the next subsection discusses which parameters to actually use for FSR in order to get a high reconstruction quality. Afterwards, an evaluation of the reconstruction quality based on a large test data set follows. All the simulations have been conducted by taking the reference images and subsampling them at non-regular positions. This way, the reconstructed image can be compared to the original image and an objective evaluation of the reconstruction performance is possible.

\begin{table}
	\caption{Model generation parameters for FSR.}
	\begin{center}
		\begin{tabular}{|l|c|}
			\hline Block size  & $4\times 4$ \\ 
			\hline Border width & $14$ \\ 
			\hline FFT size & $32\times 32$  \\ 
			\hline Iterations & $100$ \\ 
			\hline Decay factor $\hat{\rho}$ & $0.7$  \\ 
			\hline Orthogonality deficiency compensation $\gamma$ & $0.5$ \\ 
			\hline Weighting of already reconstructed areas $\delta$ & $0.5$ \\ 
			\hline 
		\end{tabular} 
		\label{tab:reconstruction_parameters}
	\end{center}
\end{table}

\subsection{Reconstruction Parameters} \label{ssec:reconstruction_parameters}

Regarding the modeling process described in the previous section, it becomes apparent that FSR relies on several different parameters. As it has to be avoided that the parameters of the final evaluation are fitted to the underlying data set, the determination of a good parameter set has to be conducted on a data set which is independent from the actual test data set. Thus, the Kodak test data base \cite{Kodak_Test_Data} has been used for determining the parameters, while the final results that are presented in the next subsection are performed on the TECNICK image data base \cite{Asuni2011}.

Accordingly, in order to identify a good parameter set for the reconstruction, a large number of different parameter combinations have been evaluated on the images from the Kodak test data base. In this context, only the luminance component has been considered. Furthermore, three different subsampling densities have been evaluated. In the most challenging case, only $10\%$ of the samples from the original image are available. The other two cases consist of $25\%$ and $50\%$ of the samples. Apparently, due to the number of involved parameters, a full search of the parameter space is not feasible. Thus, the parameters that have been proposed in earlier publications for error concealment and resolution enhancement have been used as starting points. The metric that has been used for identifying a promising parameter set is the Peak Signal-to-Noise Ratio (PSNR) between original image and reconstructed image.

In order to keep the computational load manageable, the transform size has been set to $32\times 32$ at the beginning. As this also defines the maximum possible size of the reconstruction area $\mathcal{L}$, the sum of the block size and two times the border width also has to be not larger than $32$. However, as tests with larger block sizes and according transform sizes have shown, this is no big limitation as the reconstruction quality decreases with larger block sizes, again \cite{Seiler2011c}. Taking the constraint for the block size and the border width into account, the remaining parameters have been varied in wide ranges for identifying a good set. 

As output from the parameter training, it can be stated that a block size of $4$ samples and a border width of $14$ samples is suited well for a good reconstruction quality. Furthermore, the number of iterations necessary for the model generation should be set to $100$ and the weighting function should decay with $\hat{\rho}=0.7$. Samples that already have been reconstructed before should further be weighted by $\delta=0.5$ and the orthogonality deficiency compensation factor $\gamma$ should also be set to $0.5$. To give a compact overview of the parameters, they are also listed in Table \ref{tab:reconstruction_parameters}.

In order to check how sensitive these parameters are, \mbox{Figure \ref{fig:kodak_psnr_over_parameters}} exemplarily shows the average reconstruction quality for the whole Kodak test data base with varying weighting function decay factors $\hat{\rho}$, orthogonality deficiency compensation factors $\gamma$, for different numbers of iterations to be carried out, and varying weighting factors $\delta$ of already reconstructed areas. For the plots, only the shown parameters have been changed, while all other parameters are selected according to Table \ref{tab:reconstruction_parameters}. It can be observed that the sensitivity of the parameters is not very high and they do not have to be tuned to a specific value in order to allow a high reconstruction quality. Instead, they can be varied over a relatively wide value range without having a large impact on the average reconstruction quality. 

Regarding the plots in Figure \ref{fig:kodak_psnr_over_parameters}, it can be observed that especially the sensitivity of the parameter $\delta$ which controls the reuse of already reconstructed samples is low. Nevertheless, it is of high importance as if it was set to zero no reconstruction would be possible for the case that the reconstruction area did not contain just a single originally known sample. This might happen for example for very low subsampling densities or if the available samples are unequally distributed. It can be seen quite well, that the impact of the parameters differs for the considered subsampling densities. However, in order to determine a parameter set that is universal, the proposed parameters in Table \ref{tab:reconstruction_parameters} provide a good compromise as they achieve a good reconstruction quality for the different regarded subsampling densities.

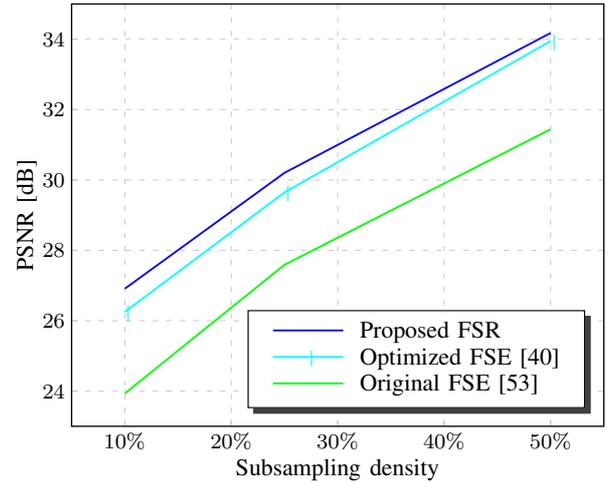
\begin{figure}
	\centering



\providelength{\AxesLineWidth}       \setlength{\AxesLineWidth}{0.5pt}
\providelength{\GridLineWidth}       \setlength{\GridLineWidth}{0.4pt}
\providelength{\GridLineDotSep}      \setlength{\GridLineDotSep}{0.4pt}
\providelength{\MinorGridLineWidth}  \setlength{\MinorGridLineWidth}{0.4pt}
\providelength{\MinorGridLineDotSep} \setlength{\MinorGridLineDotSep}{0.8pt}
\providelength{\plotwidth}           \setlength{\plotwidth}{7cm} 
\providelength{\LineWidth}           \setlength{\LineWidth}{0.7pt}
\providelength{\MarkerSize}          \setlength{\MarkerSize}{4pt}
\newrgbcolor{GridColor}{0.8 0.8 0.8}

\psset{xunit=2.000000\plotwidth,yunit=0.066667\plotwidth}
\begin{pspicture}(-0.003571,21.392857)(0.550000,35.000000)


\psline[linestyle=dashed,dash=2pt 3pt,dotsep=\GridLineDotSep,linewidth=\GridLineWidth,linecolor=GridColor](0.100000,23.000000)(0.100000,35.000000)
\psline[linestyle=dashed,dash=2pt 3pt,dotsep=\GridLineDotSep,linewidth=\GridLineWidth,linecolor=GridColor](0.200000,23.000000)(0.200000,35.000000)
\psline[linestyle=dashed,dash=2pt 3pt,dotsep=\GridLineDotSep,linewidth=\GridLineWidth,linecolor=GridColor](0.300000,23.000000)(0.300000,35.000000)
\psline[linestyle=dashed,dash=2pt 3pt,dotsep=\GridLineDotSep,linewidth=\GridLineWidth,linecolor=GridColor](0.400000,23.000000)(0.400000,35.000000)
\psline[linestyle=dashed,dash=2pt 3pt,dotsep=\GridLineDotSep,linewidth=\GridLineWidth,linecolor=GridColor](0.500000,23.000000)(0.500000,35.000000)
\psline[linestyle=dashed,dash=2pt 3pt,dotsep=\GridLineDotSep,linewidth=\GridLineWidth,linecolor=GridColor](0.050000,24.000000)(0.550000,24.000000)
\psline[linestyle=dashed,dash=2pt 3pt,dotsep=\GridLineDotSep,linewidth=\GridLineWidth,linecolor=GridColor](0.050000,26.000000)(0.550000,26.000000)
\psline[linestyle=dashed,dash=2pt 3pt,dotsep=\GridLineDotSep,linewidth=\GridLineWidth,linecolor=GridColor](0.050000,28.000000)(0.550000,28.000000)
\psline[linestyle=dashed,dash=2pt 3pt,dotsep=\GridLineDotSep,linewidth=\GridLineWidth,linecolor=GridColor](0.050000,30.000000)(0.550000,30.000000)
\psline[linestyle=dashed,dash=2pt 3pt,dotsep=\GridLineDotSep,linewidth=\GridLineWidth,linecolor=GridColor](0.050000,32.000000)(0.550000,32.000000)
\psline[linestyle=dashed,dash=2pt 3pt,dotsep=\GridLineDotSep,linewidth=\GridLineWidth,linecolor=GridColor](0.050000,34.000000)(0.550000,34.000000)

\psline[linewidth=\AxesLineWidth,linecolor=GridColor](0.100000,23.000000)(0.100000,23.180000)
\psline[linewidth=\AxesLineWidth,linecolor=GridColor](0.200000,23.000000)(0.200000,23.180000)
\psline[linewidth=\AxesLineWidth,linecolor=GridColor](0.300000,23.000000)(0.300000,23.180000)
\psline[linewidth=\AxesLineWidth,linecolor=GridColor](0.400000,23.000000)(0.400000,23.180000)
\psline[linewidth=\AxesLineWidth,linecolor=GridColor](0.500000,23.000000)(0.500000,23.180000)
\psline[linewidth=\AxesLineWidth,linecolor=GridColor](0.050000,24.000000)(0.056000,24.000000)
\psline[linewidth=\AxesLineWidth,linecolor=GridColor](0.050000,26.000000)(0.056000,26.000000)
\psline[linewidth=\AxesLineWidth,linecolor=GridColor](0.050000,28.000000)(0.056000,28.000000)
\psline[linewidth=\AxesLineWidth,linecolor=GridColor](0.050000,30.000000)(0.056000,30.000000)
\psline[linewidth=\AxesLineWidth,linecolor=GridColor](0.050000,32.000000)(0.056000,32.000000)
\psline[linewidth=\AxesLineWidth,linecolor=GridColor](0.050000,34.000000)(0.056000,34.000000)

{ \footnotesize 
\rput[t](0.100000,22.820000){$10\%$}
\rput[t](0.200000,22.820000){$20\%$}
\rput[t](0.300000,22.820000){$30\%$}
\rput[t](0.400000,22.820000){$40\%$}
\rput[t](0.500000,22.820000){$50\%$}
\rput[r](0.044000,24.000000){$24$}
\rput[r](0.044000,26.000000){$26$}
\rput[r](0.044000,28.000000){$28$}
\rput[r](0.044000,30.000000){$30$}
\rput[r](0.044000,32.000000){$32$}
\rput[r](0.044000,34.000000){$34$}
} 

\pspolygon[linewidth=\AxesLineWidth](0.050000,23.000000)(0.550000,23.000000)(0.550000,35.000000)(0.050000,35.000000)(0.050000,23.000000)

{ \small 
\rput[b](0.300000,21.392857){
\begin{tabular}{c}
Subsampling density\\
\end{tabular}
}

\rput[t]{90}(-0.003571,29.000000){
\begin{tabular}{c}
PSNR [dB]\\
\end{tabular}
}
} 

\newrgbcolor{color252.0022}{0  0  1}
\savedata{\mydata}[{
{0.100000,26.909037},{0.250000,30.201875},{0.500000,34.175686}
}]
\dataplot[plotstyle=line,showpoints=true,dotstyle=+,dotsize=\MarkerSize,linestyle=solid,linewidth=\LineWidth,linecolor=color252.0022]{\mydata}

\newrgbcolor{color253.0016}{0  1  0}
\savedata{\mydata}[{
{0.100000,23.930422},{0.250000,27.586712},{0.500000,31.439044}
}]
\dataplot[plotstyle=line,showpoints=true,dotstyle=asterisk,dotsize=\MarkerSize,linestyle=solid,linewidth=\LineWidth,linecolor=color253.0016]{\mydata}

\newrgbcolor{color254.0015}{0  1  1}
\savedata{\mydata}[{
{0.100000,26.248507},{0.250000,29.643526},{0.500000,33.944557}
}]
\dataplot[plotstyle=line,showpoints=true,dotstyle=|,dotsize=\MarkerSize,linestyle=solid,linewidth=\LineWidth,linecolor=color254.0015]{\mydata}

{ \small 
\rput[br](0.538000,23.360000){%
\psshadowbox[framesep=0pt,linewidth=\AxesLineWidth]{\psframebox*{\begin{tabular}{l}
\Rnode{a1}{\hspace*{0.0ex}} \hspace*{0.7cm} \Rnode{a2}{~~Proposed FSR} \\
\Rnode{a5}{\hspace*{0.0ex}} \hspace*{0.7cm} \Rnode{a6}{~~Optimized FSE \cite{Seiler2011c}} \\
\Rnode{a3}{\hspace*{0.0ex}} \hspace*{0.7cm} \Rnode{a4}{~~Original FSE \cite{Meisinger2004a}} \\
\end{tabular}}
\ncline[linestyle=solid,linewidth=\LineWidth,linecolor=color252.0022]{a1}{a2} \ncput{\psdot[dotstyle=+,dotsize=\MarkerSize,linecolor=color252.0022]}
\ncline[linestyle=solid,linewidth=\LineWidth,linecolor=color253.0016]{a3}{a4} \ncput{\psdot[dotstyle=asterisk,dotsize=\MarkerSize,linecolor=color253.0016]}
\ncline[linestyle=solid,linewidth=\LineWidth,linecolor=color254.0015]{a5}{a6} \ncput{\psdot[dotstyle=|,dotsize=\MarkerSize,linecolor=color254.0015]}
}%
}%
} 

\end{pspicture}%
	\caption{Average reconstruction quality in PSNR with respect to different subsampling densities for Kodak test data base for FSR and two of its ancestors. }
		\label{fig:psnr_over_density_fse_evolution}
\end{figure}

As mentioned above, the proposed FSR is based on an older algorithm which is called Frequency Selective Extrapolation (FSE) which also already has an evolution of several years. In order to assess the progress of this algorithm, Figure \ref{fig:psnr_over_density_fse_evolution} shows the reconstruction quality for different subsampling densities for three different stages of development: First, this is the original version of FSE \cite{Meisinger2004a}, without the weighting function, the orthogonality deficiency compensation, and the above introduced enhancements. Hence, it can be regarded as a direct Matching Pursuits \cite{Mallat1993} based modeling of the available samples. The second algorithm is a highly optimized version of FSE \cite{Seiler2011c} which already includes the spatial weighting function, orthogonality deficiency compensation and a processing order optimized for large loss areas. And finally, the plot shows the results for the proposed FSR which includes the processing order suited for the considered reconstruction task and the frequency weighting. While the gain of FSR is moderate for large subsampling densities where many of the samples are still available, it clearly increases for low densities and reaches up to $0.7 \dB$ over the already optimized FSE \cite{Seiler2011c}. 

\begin{table*}
	\caption{Average gain (with standard deviation) in $\dB$ PSNR compared to Linear Interpolation. Maximum gain written in bold font.}
	\begin{center} 
		\setlength{\tabcolsep}{1.9mm}
		\begin{tabular}{|l|c|c|c|c|c|c|c|c|}
			\hline Subsampling  &  FSR  &  NNI \cite{Sibson1981}  &  BL \cite{Papoulis1975, Gerchberg1974}  &  KR \cite{Takeda2007}  &  MCA \cite{Elad2005} &  TV \cite{Dahl2010}  &  WI \cite{Starck2010} & CLS \cite{Afonso2011} \\
			density &&&&&&&&\\
			\hline $10\%$ & $\bf{1.45} \pm 0.99$ & $0.24 \pm 0.04$ & $-4.44 \pm 0.69$ & $-0.14 \pm 0.50$ & $-1.16 \pm 1.39$ & $-1.84 \pm 0.97$ & $-1.09 \pm 0.68$ & $-1.32 \pm 0.76$ \\
			\hline $25\%$ & $\bf{2.17} \pm 1.26$ & $0.25 \pm 0.05$ & $-5.94 \pm 0.90$ & $0.90 \pm 1.09$ & $-1.07 \pm 1.58$ & $-1.03 \pm 0.88$ & $0.47 \pm 0.91$ & $-0.70 \pm 0.81$ \\
			\hline $50\%$ & $\bf{2.41} \pm 1.33$ & $0.38 \pm 0.06$ & $-7.40 \pm 1.19$ & $-0.24 \pm 1.27$ & $-0.88 \pm 1.75$ & $-0.04 \pm 0.97$ & $1.63 \pm 1.36$ & $0.39 \pm 0.98$ \\
			\hline $75\%$ & $2.39 \pm 1.35$ & $0.68 \pm 0.14$ & $-9.04 \pm 1.38$ & $-1.00 \pm 1.39$ & $-0.59 \pm 1.93$ & $0.96 \pm 1.09$ & $\bf{2.45} \pm 1.85$ & $1.66 \pm 1.18$ \\
			\hline $90\%$ & $2.47 \pm 1.36$ & $1.13 \pm 0.26$ & $-10.84 \pm 1.46$ & $-1.11 \pm 1.39$ & $-0.28 \pm 2.00$ & $1.65 \pm 1.12$ & $\bf{2.95} \pm 2.01$ & $2.63 \pm 1.38$ \\
			\hline 
		\end{tabular}
		\label{tab:average_gain_over_linear}
	\end{center}
\end{table*}

\subsection{Evaluation of the Reconstruction Quality} \label{ssec:reconstruction_quality}

After having derived a useful set of parameters in the preceding subsection, the actual abilities of FSR for the resampling of images to a regular grid are evaluated in the following. For this, the simulations have been carried out on the TECNICK  image data base \cite{Asuni2011} which is independent from the Kodak data base that has been used for determining the parameter set. The data base comprises $100$ images which are of size $1200\times 1200$ pixels and again only the luminance component of the images is considered. The individual images from this data base contain very different content and therewith challenge the reconstruction in very different ways.

\begin{figure}
	\centering
	\input{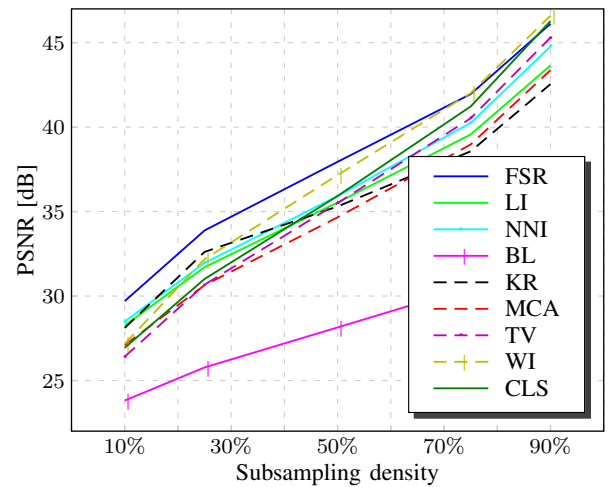}
	\caption{Reconstruction quality in PSNR for different subsampling densities, averaged over whole TECNICK image data base \cite{Asuni2011}.}
	\label{fig:psnr_over_density_testimages}
\end{figure}

For a comprehensive evaluation of the performance, FSR furthermore is compared to several algorithms that can also be used for this reconstruction task. The first one is Linear Interpolation (LI) where the reconstruction of an unknown sample is achieved by determining the three closest neighbors and linearly interpolating between them. Additionally, Natural Neighbor Interpolation (NNI) \cite{Sibson1981} is considered as well as a Band-limited Reconstruction (BL) based on the concepts from \cite{Papoulis1975, Gerchberg1974}. Besides these algorithms, the statistically driven Steering Kernel Regression (KR) \cite{Takeda2007} is evaluated. Furthermore, a reconstruction making use of Total Variation Minimization (TV) \cite{Dahl2010} is considered. Since the proposed FSR makes use of a sparse modeling, two alternative algorithms that exploit this property are considered, as well. First, this is the reconstruction using \mbox{Morphological Component Analysis (MCA) \cite{Elad2005}.} This algorithms decomposes an image into a texture and a cartoon layer and performs a sparse modeling of these by making use of Basis Pursuits Denoising \cite{Chen1998} and a Wavelet representation. Second, by testing the sparsity-based Wavelet Inpainting (WI) \cite{Starck2010}, an alternative algorithm is considered that also exploits the sparsity property of image signals. For this, WI performs an iterative hard thresholding of Wavelet coefficients and aims at finding a sparse representation of the signal in the Wavelet domain. As Wavelets are able to directly handle the instationarity of image signals, a block-wise processing as for the proposed FSR is not necessary, there. And finally, Constrained Split Augmented Lagrangian Shrinkage Algorithm (CLS) \cite{Afonso2011} is considered as another modern regularization minimization algorithm for solving this inverse problem.

The simulations for evaluating the reconstruction performance have been carried out for seven different subsampling densities. That is to say, from the original image non-regular subsets containing between $10\%$ and $90\%$ of the samples are taken and the objective always is to reconstruct the original image from these samples as good as possible. For every subsampling density, the sampling masks are fixed, that is to say, every reconstruction algorithm performs the reconstruction on the identical non-regularly subsampled image. The output of the simulations is shown in Figure \ref{fig:psnr_over_density_testimages} where the reconstruction quality is plotted over the subsampling density. For this, the mean reconstruction quality averaged over all images from the TECNICK data base is considered. It can be seen quite well that FSR outperforms all the other algorithms considerably for low subsampling densities where only very few of the original samples are available. For these very challenging densities, gains of more than $1 \dB$ PSNR over the second best algorithm are possible. Only for the case where most of the samples from the original image are still available, FSR is beaten by WI and CLS. However, it has to be noted that the FSR parameter set is fixed and not adapted to the subsampling density. As has been shown in the previous subsection, it might be possible to achieve a higher reconstruction quality by choosing the parameters to fit higher densities, but as all other algorithms are also not adapted to the subsampling density, this option is not considered. 

As the used test data base consists of $100$ images, it is also possible to analyze the statistics of the reconstruction quality. For this, Table \ref{tab:average_gain_over_linear} lists the average gain of the different algorithms compared to Linear Interpolation, together with the standard deviation. While the average gain reflects the behavior from Figure \ref{fig:psnr_over_density_testimages}, the standard deviation shows how reliable these gains are. Accordingly, the lower the standard deviation is, the lower is the probability that the reconstruction of few of the test images is poor while on average the reconstruction quality might still be high. Regarding the results for FSR, it can be seen that the average gain always is considerably larger than the standard deviation, making the illustrated gains also quite reliable. Only for subsampling densities of $75\%$ and $90\%$, WI is able to slightly outperform FSR, however, the gains have a larger standard deviation, showing the quality of reconstructed images varies more for WI than for FSR. For a density of $90\%$, where only a very small number of samples is missing, CLS is able to outperform FSR, as well, and also exhibits a similar standard deviation.

\begin{table}
	\caption{Reconstruction quality in terms of SSIM. Maximum value written in bold font.}
	\begin{center}
		\begin{tabular}{|l|c|c|c|c|c|}
			\hline Subsampling  & $10\%$ & $25\%$ & $50\%$ & $75\%$ & $90\%$ \\
			density &&&&&\\
			\hline FSR                                    & $\bf{0.810}$ & $\bf{0.878}$ & $\bf{0.914}$ & $\bf{0.931}$ & $\bf{0.939}$ \\
			\hline LI                                     & $0.786$ & $0.857$ & $0.904$ & $0.927$ & $0.937$ \\
			\hline NNI \cite{Sibson1981}                  & $0.790$ & $0.861$ & $0.907$ & $0.929$ & $0.938$ \\
			\hline BL \cite{Papoulis1975, Gerchberg1974}  & $0.604$ & $0.656$ & $0.717$ & $0.771$ & $0.819$ \\
			\hline KR \cite{Takeda2007}                   & $0.792$ & $0.858$ & $0.895$ & $0.921$ & $0.934$ \\
			\hline MCA \cite{Elad2005}                    & $0.733$ & $0.818$ & $0.882$ & $0.916$ & $0.933$ \\
			\hline TV \cite{Dahl2010}                     & $0.740$ & $0.837$ & $0.901$ & $0.929$ & $0.938$ \\
			\hline WI \cite{Starck2010}                   & $0.755$ & $0.861$ & $0.911$ & $\bf{0.931}$ & $\bf{0.939}$ \\
			\hline CLS \cite{Afonso2011}                  & $0.750$ & $0.844$ & $0.905$ & $0.930$ & $\bf{0.939}$ \\
			\hline
		\end{tabular}
		\label{tab:average_ssim}
	\end{center}
\end{table}

Since PSNR is a metric just relying on the mean squared error of the reconstructed signal, it sometimes does not represent well the perceptual quality. A metric which accounts more for this is the Structural Similarity Measure (SSIM) \cite{Wang2004}. In Table \ref{tab:average_ssim} the average SSIM results are given for the considered sequences and the evaluated subsampling densities. Apparently, for the considered scenario, the SSIM results conform well with the PSNR and it can be seen that the proposed FSR is able to outperform the other algorithms for low densities significantly and for high subsampling densities provides a similar reconstruction quality as WI and CLS.

\begin{figure}
	\centering
	\input{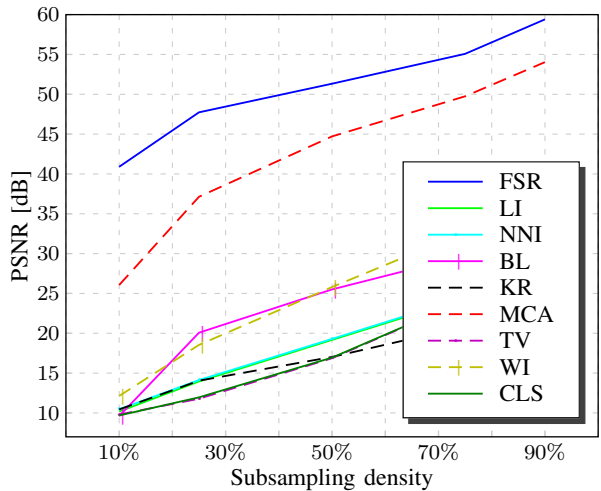}
	\caption{Reconstruction quality in PSNR for test image Zoneplate.}
	\label{fig:psnr_over_density_zoneplate}
\end{figure}

\begin{figure*}
	\centering
	\psfrag{Original}[l][l]{\color{white}Original}
	\psfrag{FSR}[l][l]{\color{white}FSR}
	\psfrag{LI}[l][l]{\color{white}LI}
	\psfrag{NNI}[l][l]{\color{white}NNI}
	\psfrag{BL}[l][l]{\color{white}BL}
	\psfrag{Sampled}[l][l]{\color{white}Subsampled}
	\psfrag{KR}[l][l]{\color{white}KR}
	\psfrag{MCA}[l][l]{\color{white}MCA}
	\psfrag{TV}[l][l]{\color{white}TV}
	\psfrag{WI}[l][l]{\color{white}WI}
	\psfrag{CS}[l][l]{\color{white}CLS}
	\includegraphics[width=0.95\textwidth]{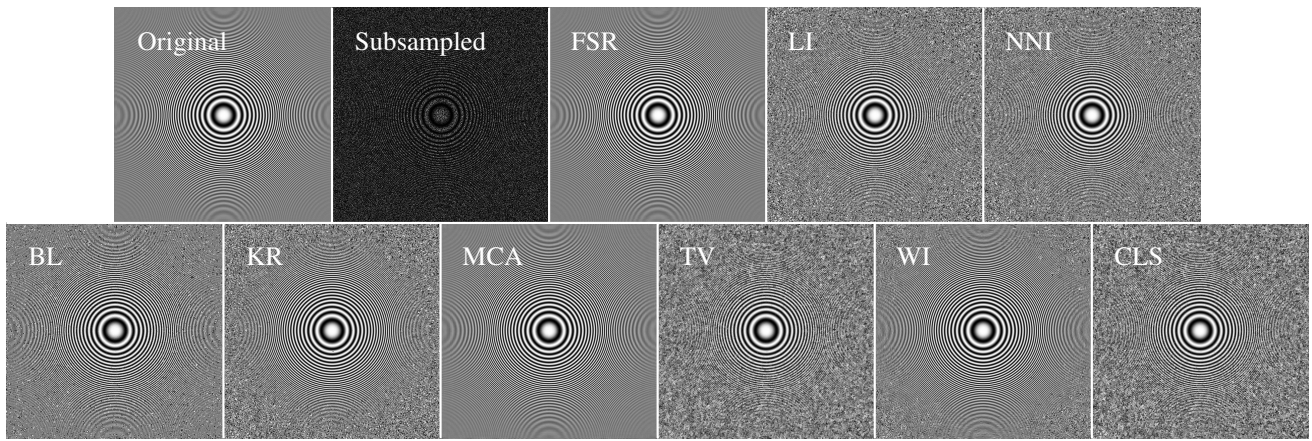}
	\caption{Visual results for test image Zoneplate with non-regular subsampling of density $25\%$ and reconstruction with different algorithms. \emph{(Please pay attention, additional aliasing may be caused by printing or scaling. Best to be viewed enlarged on a monitor.)
		} }
	\label{fig:visual_results_zoneplate}
\end{figure*}

The results presented so far show that the reconstruction method BL falls far behind all the other algorithms and only achieves a very low quality. However, it has to be noted that this algorithm reconstructs a band-limited solution given the available samples. That is to say, if for example the subsampling density $10\%$ was considered, a signal is reconstructed that also is limited to only $10\%$ of the original frequency range. Apparently, this leads to strong distortions and only a low reconstruction quality. In order to test the abilities of the different algorithms for reconstructing high frequency content, all algorithms have also been applied on the test image Zoneplate which is a rotation-symmetrically chirp and is shown top left in Figure \ref{fig:visual_results_zoneplate}. Since this image contains the whole range from very low to very high spatial frequencies, it is well suited for evaluating to what extent an algorithm can recover high frequency content. In Figures \ref{fig:psnr_over_density_zoneplate} and \ref{fig:visual_results_zoneplate}, the reconstruction of the test image Zoneplate from samples located at non-regular positions is given. While the first of the two figures shows the reconstruction quality in PSNR for the considered algorithms over varying subsampling densities, the latter shows the visual output of the reconstruction at a subsampling density of $25\%$. It can be seen quite well that FSR is able to outperform all other algorithms significantly for this test image and achieves an extremely high reconstruction quality. Especially if the reconstruction is compared to the original image in \mbox{Figure \ref{fig:visual_results_zoneplate}}, almost no difference is visible. Regarding the quality of the alternative reconstruction algorithms, it can be seen that, except for MCA, they are not able to recover high frequency content. Thus, they are only able to achieve low PSNR values and fall behind.

\begin{figure*}
\centering
\psfrag{Original}[c][l][1][90]{Original}
\psfrag{FSR}[c][l][1][90]{FSR}
\psfrag{LI}[c][l][1][90]{LI}
\psfrag{NNI}[c][l][1][90]{NNI}
\psfrag{BL}[c][l][1][90]{BL}
\psfrag{Sampled}[c][l][1][90]{Subsampled}
\psfrag{KR}[c][l][1][90]{KR}
\psfrag{MCA}[c][l][1][90]{MCA}
\psfrag{TV}[c][l][1][90]{TV}
\psfrag{WI}[c][l][1][90]{WI}
\psfrag{CS}[c][l][1][90]{CLS}
\includegraphics[width=0.96\textwidth]{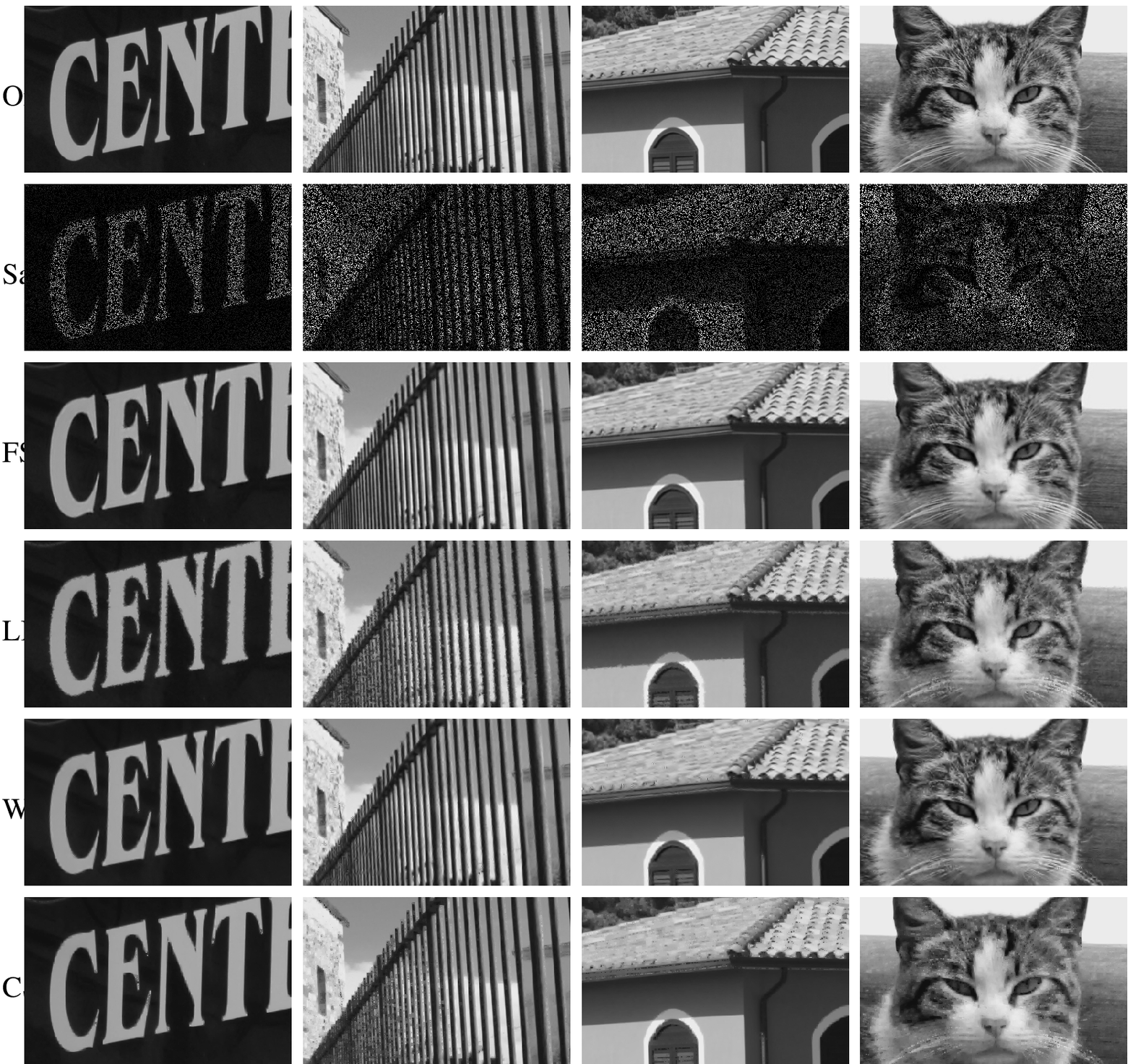}
\caption{Visual results for details of different test images from the TECNICK image data base with non-regular subsampling of density $25\%$ and reconstruction with different algorithms. \emph{(Please pay attention, additional aliasing may be caused by printing or scaling. Best to be viewed enlarged on a monitor.)
}
}
\label{fig:visual_results}
\end{figure*}

Besides the evaluation of the reconstruction quality in terms of PSNR and SSIM, of course, the visual quality of the reconstructed images is of great importance. In order to illustrate that the measured performance also is visually noticeable, Figure \ref{fig:visual_results} shows the output of the reconstruction for details of four test images from the TECNICK data base for a subsampling density of $25\%$. The shown patches have been selected to represent very different content and it can be seen that the considered algorithms behave differently with respect to the content to be reconstructed. In order to show the images in a sufficient size, only a subset of algorithms has been selected for this comparison. Besides FSR, this is LI as a rather simple method, WI as a different sparsity-based reconstruction algorithm, and CLS as an regularization minimization algorithm. Regarding the algorithms used for comparison, it can be seen that they all produce different artifacts. For example, LI produces very jagged edges and a lot of noise-like artifacts. In contrast to this, CLS is able to generate sharp edges, but in some cases leads to an oversmoothing and fine structures like the corners of the letters or the fence could not be recovered. WI achieves a very high subjective quality, however there are for example some small artifacts like impulse-noise on the roof and especially very fine structures as at the far end of the fence or the hair of the cat looks jagged, as well. In contrast to this, the proposed FSR produces a high reconstruction quality that is visually consistent. This holds for smooth as well as textured areas and even very fine structures, as at the fence, as well as sharp edges at the text letters can be recovered with high quality. Even for noise-like structures as shown in the rightmost column, a reasonable reconstruction is possible and the fur of the cat looks quite similar to the original image. Therewith, the results from the objective evaluation also reflect the visual quality quite well. 

Comparing the results presented above, it can be discovered, that FSR is able to outperform the other algorithms especially for small subsampling ratios, that is to say, when there are only few samples available for the reconstruction. This can be explained by several reasons. First, the iterative modeling in combination with the novel frequency weighting allows for a stable and high quality reconstruction of individual blocks. Second, the block-wise processing assures a high reconstruction quality on a large scale. Unlike algorithms like TV, CLS, MCA, or WI, the processing of FSR is carried out for overlapping blocks. In doing so, it can account effectively for the instationarity of image signals. By applying a block-wise Fourier-transform, a high frequency resolution can be achieved, even for small areas, allowing for the reconstruction of high-frequency content. At the same time, the optimized processing order in combination with the reuse of already reconstructed samples from one block for the processing of the next one is important for achieving a high quality. As the local density of the samples determines the processing order, it can be assured that regions with many available samples are processed prior to regions with only few available samples. This is important in combination with the reuse of already reconstructed pixels as it is possible then to extend the content from already reconstructed blocks into areas where only very few samples are available. This may happen especially in the case of very small subsampling densities where only few samples are available. In this case, the other algorithms sometimes produce artifacts like oversmoothing as can be seen for example in Figure \ref{fig:visual_results}. For high subsampling densities, this case is less likely. Hence, CLS and WI can achieve a high quality as well, or even are able to outperform FSR.

\begin{table}
	\caption{Average processing time in seconds for one image from TECNICK image data base \cite{Asuni2011} with subsampling density $10\%$.}
	\begin{center}
		\begin{tabular}{|l|r|}
			\hline FSR  & $463.00$ \\ 
			\hline LI & $4.35$ \\ 
			\hline NNI \cite{Sibson1981} & $5.17$  \\ 
			\hline BL \cite{Papoulis1975, Gerchberg1974} & $56.10$ \\ 
			\hline KR \cite{Takeda2007} & $170.00$  \\ 
			\hline MCA \cite{Elad2005} & $18500.00$ \\ 
			\hline TV \cite{Dahl2010} & $142.00$ \\ 
			\hline WI \cite{Starck2010} & $11900.00$ \\
			\hline CLS \cite{Afonso2011} & $85.60$ \\
			\hline
		\end{tabular} 
		
		\label{tab:runtime}
	\end{center}
\end{table}

In order to assess the complexity of the different algorithms, the average time for reconstructing one image from the TECNICK image data base at a subsampling density of $10\%$ has been measured. The results are listed in Table \ref{tab:runtime}. The tests have been carried out on an Intel Xeon E5-1620 v2 running at $3.70$ GHz, equipped with $32$ GB of RAM and running MATLAB R2013b. It can be seen that the runtime of the different algorithms spreads over four magnitudes. Additionally, all the algorithms which are able to achieve a high reconstruction quality require at least several tens of seconds. And while the proposed FSR is of course not the fastest of them, with $460$ seconds for reconstructing an image of $1200\times 1200$ pixels, it still exhibits a manageable complexity. The processing time for FSR could be further reduced by using larger block sizes. For example, if the transform size was left unchanged, doubling the block size would speed up the processing by a factor of four as the number of blocks to be processed accordingly is reduced by a factor of four, in the same way. However, as shown in \cite{Seiler2011c} smaller block sizes typically lead to a higher reconstruction quality. In addition to this, by using for example a block size of $8\times 8$, either the border width would have to be reduced or a larger transform size would have to be used. While the latter one would eat up parts of the acceleration, the former would lead to a reduced reconstruction quality. Hence, the parameters given in Table \ref{tab:reconstruction_parameters} provide a good compromise for the reconstruction by FSR.


\section{Conclusion and Outlook} \label{sec:conclusion} 
In this paper, Frequency Selective Reconstruction was introduced for resampling images to a regular grid for the case that the image information only is available for a non-regular subset of pixel positions. The proposed algorithm is able to recover the image signal on a regular grid at a very high quality, even in the case that only very few samples are available. For this, the property of image signals is exploited that small areas of images can be sparsely represented in the Fourier domain. In this process, the property of imaging systems that high-frequency content is less likely to occur than low-frequency content is used for obtaining a stable estimation. 

In doing so, the proposed algorithm achieves a very high reconstruction quality and is able to outperform state-of-the-art reconstruction algorithms. This holds especially for the case that the ratio between the available non-regularly spaced pixels and the samples on the desired regular grid becomes very small. In this case, gains of more than $1 \dB$ PSNR compared to state-of-the-art algorithms are possible. These results are also representative for the visual quality of the reconstruction as the proposed algorithm is able to achieve a very high visual quality as well, which is of course of great importance for reconstruction algorithms. Furthermore, the proposed algorithm is especially suited for reconstructing high frequency content and very fine details.

The very high reconstruction quality and the ability of the algorithm to recover high-frequency content from only a small number of non-regularly spaced pixels allow for alternative sampling concepts where the non-regularity is especially exploited for increasing the sensor resolution as shown for example in \cite{Schoberl2011a}. 

Besides this, future research aims at using alternative basis functions sets for the reconstruction which may also be adaptive to the underlying signals. Furthermore, extensions to video signals are envisioned by either using a three-dimensional model generation similar to \cite{Seiler2011} or a joint spatial and temporal reconstruction similar to \cite{Seiler2008b} which would only introduce little computational overhead to the proposed two-dimensional reconstruction.



\end{document}